\documentclass[]{aa}
\usepackage{graphicx}
\usepackage{txfonts}
\usepackage{natbib}
\usepackage{mathrsfs}
\usepackage{longtable}
\usepackage[shortlabels]{enumitem}
\usepackage{siunitx}
\usepackage{epstopdf}
\usepackage{hyperref}	
\hypersetup{colorlinks=true,linkcolor=blue,citecolor=blue,filecolor=blue,urlcolor=blue}
\usepackage{comment}
\usepackage{txfonts}
\usepackage{url}
\usepackage{stfloats}

\usepackage[dvipsnames]{xcolor}
\usepackage{overpic}
\usepackage{orcidlink}

\newcommand{\gaia}{{\it Gaia }}

\usepackage{academicons}
\definecolor{orcidlogocol}{HTML}{A6CE39}

\begin{document}
\title{Beyond the two-infall model} \subtitle{I.  Indications for  a recent  gas infall    with \gaia DR3 chemical abundances }

  \author { E. Spitoni \orcidlink{0000-0001-9715-5727} \inst{1}  \thanks {Email to: emanuele.spitoni@oca.eu},
  A. Recio-Blanco  \orcidlink{0000-0002-6550-7377} \inst{1}, P. de Laverny \orcidlink{0000-0002-2817-4104}\inst{1},   P. A. Palicio    \orcidlink{0000-0002-7432-8709} \inst{1} , G. Kordopatis \orcidlink{0000-0002-9035-3920}\inst{1},\\  M. Schultheis \orcidlink{0000-0002-6590-1657} \inst{1}, G. Contursi \orcidlink{0000-0001-5370-1511} \inst{1}, E. Poggio \orcidlink{0000-0003-3793-8505}\inst{1,2}, D. Romano   \orcidlink{0000-0002-0845-6171} \inst{3}, F. Matteucci \orcidlink{0000-0001-7067-2302}\inst{4,5,6} }
  \institute{Universit\'e C\^ote d'Azur, Observatoire de la C\^ote d'Azur, CNRS, Laboratoire Lagrange, Bd de l'Observatoire,  CS 34229, 06304 Nice cedex 4, France  
    \and Osservatorio Astrofisico di Torino, Istituto Nazionale di Astrofisica (INAF), I-10025 Pino Torinese, Italy
  \and INAF, Osservatorio di Astrofisica e Scienza dello Spazio, Via Gobetti 93/3, 40129 Bologna, Italy
  \and Dipartimento di Fisica, Sezione di Astronomia, Università di Trieste, via G.B. Tiepolo 11, I-34131, Trieste, Italy \and I.N.A.F. Osservatorio Astronomico di Trieste, via G.B. Tiepolo 11, I-34131, Trieste, Italy \and I.N.F.N. Sezione di Trieste, via Valerio 2, 34134 Trieste, Italy
}

\date{Received xxxx / Accepted xxxx}

\abstract
{The recent Gaia Data Release 3 (DR3) represents an unparalleled revolution in Galactic Archaeology, providing us with numerous radial velocities chemical abundances for millions of stars  with all-sky coverage.}
{We present a new chemical evolution model for the Galactic disc components  (high- and low- $\alpha$ sequence stars) designed to reproduce the new abundance ratios provided by  the GSP-spec module for the  \gaia DR3   and also constrained  by  the detailed star formation (SF) histories for both the thick and thin
disc stars inferred from previous \gaia releases. }
{   Sophisticated modeling based on previous  \gaia releases have found evidence for   narrow episodes of
enhanced SF inferred in recent time.  Additionally, \gaia DR3  indicated the presence of young (massive)  low-$\alpha$ disc stars  which show evidence of a recent chemical impoverishment in several elements. In order  to reproduce these observables, we propose a new  chemical evolution model in which  the low-$\alpha$ sequence  is generated by two distinct infall episodes. Hence, in this study we compare \gaia DR3 chemical abundances with the predictions of a {\it three-infall} chemical evolution model for the high- and low-$\alpha$ components.
}
{The proposed {\it three-infall} chemical evolution model nicely reproduces  the main features of the abundance ratio [X/Fe] versus [M/H] (X=Mg, Si, Ca, Ti, $\alpha$) of \gaia DR3 stars  in different age bins for the considered $\alpha$ elements. Moreover, the most recent gas  infall - which started  $\sim$ 2.7 Gyr ago -  allows us to predict well the \gaia DR3 young population  which has experienced a recent chemical impoverishment.
}
 {We extended previous chemical evolution models designed to reproduce APOGEE and APOKASC data to predict new \gaia DR3 chemical abundances. To this aim we proposed a {\it three-infall} chemical evolution model in order to trace better both i) the young population in  \gaia DR3 with evidence of chemical impoverishment and  ii) the SF history from previous \gaia releases.  }


\titlerunning{Beyond the two-infall model with \gaia DR3}

\authorrunning{Spitoni et al.}

\maketitle
                                                    \section{Introduction}

 Unravelling the origin and evolution of   our Galaxy's disc ultimately relies on the  study and interpretation of signatures
imprinted in resolved stellar populations, such as their chemical and kinematic properties \citep{freeman2002}. 
Intermediate releases of   \gaia   mission \citep{gaia2016, gaia2_2018} have already   dramatically improved the  comprehension  of  the structure and evolution of our Milky Way mainly from their unprecedentedly   astrometric and line-of-sight velocity measurements \citep{antoja2021,laporte2019}.

However,  Galactic Archaeology relies also on the interpretations of chemical   signatures present  
in the stellar atmospheres \citep{matteucci2021,matteucci2012,freeman2002,feltzing2016}. In fact, once a star is born,   the chemical enrichment history of the interstellar medium (ISM) from which it was formed is imprinted in its atmosphere.  For this
reason,
in the stellar atmospheres we can find insights of the processes that determined the formation and regulated the evolution of the various components of our Galaxy.
Hence, previous  studies focused on the   Galactic Archaeology had to complement  the intermediate \gaia data releases with chemical data from ground-based observations. However,  ground-based surveys like the Galactic Archaeology with HERMES survey  \citep[GALAH;][]{buder2021}, the Apache Point Observatory Galactic Evolution Experiment project \citep[APOGEE;][]{Majewski:2017ip,Ahumada2019,apogeedr172022}, Gaia-ESO Survey \citep{randich2022}, The Radial Velocity Experiment \citep[RAVE;][]{steinmetz2020} and Large sky Area Multi Object fiber Spectroscopic Telescope \citep[LAMOST;][]{zheng2021} carried  on all the issues of biased samples  which hampered the observations from Earth.
In this context,  \gaia Data Release 3 (DR3; \citealt{vallenari2022}) and  \citet{recioDR32022a,recioDR32022b} have brought a truly and unprecedented revolution opening a new
era of all-sky spectroscopy. 
With about 5.6 million stars, the \gaia DR3 General Stellar Parametrizer - spectroscopy   (GSP-Spec, \citealt{recioDR32022a}) all-sky catalogue is the largest compilation of stellar chemo-physical parameters and the first one from space data.

The analysis of  spectroscopic data from ground-based surveys such as  the APOGEE  \citep[e.g.][]{hayden2015,queiroz2020}, the Gaia-ESO   \citep[e.g.][]{RecioBlanco:2014dd,kordopatis2015,RojasArriagada:2016eq,RojasArriagada:2017ka}, the AMBRE project  \citep{Mikolaitis:2017gd,delaverny2013}, GALAH  \citep{buder2019,buder2021}  suggest the existence of a clear separation between two sequences of disc stars in the [$\alpha$/Fe] versus [Fe/H] abundance ratio space: the so-called high-$\alpha$  and low-$\alpha$ sequences.

Several theoretical models of the evolution of Galactic discs  have suggested that the  bimodality may be strictly connected to a 
delayed accretion of gas of primordial chemical composition
 (or with a metal poor chemical composition). By revising the classical two-infall chemical evolution model by \citet{chiappini1997} and \citet{grisoni2017},  \citet{spitoni2019, spitoni2020, spitoni2022}  showed that a significant delay of $\sim$ 4 Gyr between two consecutive episodes of 
gas accretion is needed to explain the dichotomy in the  APOKASC  (APOGEE+ Kepler Asteroseismology Science Consortium, \citealt{pinso2014}) sample for the solar neighbourhood  \citep{victor2018} and APOGEE DR16 stars.
In particular, they predict that the  star formation
rate (SFR) has a minimum at age of $\sim \! 8$\,Gyr. A similar quenching of star formation
(SF) at an age of  8\,Gyr was derived by \citet{snaith2015}  using the chemical abundances of 
 \citet{adi2012}  and the isochrone ages of  \citet{haywood2013} for 
solar-type stars. \citet{katz2021} concluded that in APOGEE data is clear  the signature of a dilution of the interstellar medium from 6 kpc to the outskirt of the disc, which occurred before the onset of the thin disc formation.

  By analysing ESO/HARPS spectra  of local solar twin stars, \citet{nissen2020} found that the age-metallicity distribution has two distinct populations with a clear age dissection. The authors suggest that these two sequences  may be interpreted as evidence of two episodes of accretion of gas onto the Galactic disc with  quenching of SF in between them, which is in  agreement with the scenario proposed by \citet{spitoni2019}  and \citet[][hereafter ES20]{spitoni2020}. Finally, \citet{romano2020}, comparing chemical evolution models  with the  recent  [C/Fe] and [C/O] abundance ratios from  
 high-resolution spectra by \citet{amarsi2019} for dwarf stars, concluded  that the delayed gas infall scenario proposed by \citet{spitoni2019} fits reasonably well also those data.

 The new AMBRE:HARPS data   were reproduced by \citet{palla2022} with chemical evolution models characterised by  peculiar histories of SF, such as  the two-infall model with a significant delay. 
  \citet{xiang2022}, analysing subgiant stars in  LAMOST, confirmed that the stellar age-metallicity distribution  splits into two almost disjoint parts, separated at age $\simeq$ 8 Gyr. In \citet{sahlholdt2022}, they also  highlighted  age-metallicity relation  characterised by several disconnected sequences which could be associated to different  SF regimes throughout the  Milky Way disc evolution.
    In \citet{spitoni2022}, it was finally shown  the signature of a delayed gas infall episode which gives rise to a hiatus in the SF history of the Galaxy is imprinted both in the [Mg/Fe] versus [Fe/H] relation and in vertical distribution of [Mg/Fe] abundances in the solar vicinity.

Analysing [Ca/Fe] versus [M/H] abundance ratios of \gaia DR3 disc stars, \citet{recioDR32022b}   found   that    most of 7300 massive stars of the sample (young objects selected with the  criteria indicated in their Section 3.2.2),  are Ca-poor with [Ca/Fe] values down to $\sim$-0.3 dex with [M/H] $\in$ [-0.5,+0.0] dex. Hence,  young disc stars show, surprisingly, a recent chemical impoverishment in several elements. It is worth mentioning that also in  the [Ce/Fe] versus [M/H] ratio, a metal impoverished population in the low-$\alpha$ sequence exists. The interesting fact is that  these stars are exactly the  same massive objects mentioned above in  the  [Ca/Fe] versus [M/H] sample.
Only a large number statistics chemical survey as the \gaia DR3    GSPspec one could clearly provide constraints to this younger population. 
Many young stars are relatively hot main sequence objects for which the detailed chemical abundance analysis is not possible (due to a lack of metallic lines in the spectra). On the other hand, cool young stars are rare (about 30 000 stars out of 5.5 million stars in the GSPspec catalogue as shown in \citealt{recioDR32022b}).
Even if in this paper we concentrate on the solar neighbourhood, \citealt{recioDR32022b}) shows that the impoverishment of the Massive population is apparent at all Galactic radii.

Stellar migration cannot be  invoked to explain such a young metal-impoverished  stellar population  because, due to the decreasing stellar density  profile with the Galactocentric distance, the inwards stellar migration that decreases the local  metallicities should not dominate. However,  a more likely physical process responsible for this  recent chemical impoverishment could be the {\it dilution} caused by a recent gas infall event.

In this article, we  present a new  chemical evolution model  designed to reproduce the abundance ratios for $\alpha$-elements in \gaia DR3, and in particular the young population which seems to show  recent chemical impoverishment in several elements.
In addition, our model will  be constrained  by the several brief episodes of enhanced SF  happened in recent times  (in the last 2 Gyr of disc evolution) as inferred  from
\gaia DR2-observed colour–magnitude diagrams by  \citet{lara2020}.  The authors proposed that the timing of these enhanced SF episodes is consistent  with the  Sagittarius dwarf spheroidal galaxy  pericentre passages  which  triggered the  SF in the Galactic disc \citep[e.g.,][]{laporte2019b,antoja2020}.

The paper 
is organised as follows.  In Section \ref{s:gaia}, the \gaia DR3 sample in the solar vicinity  is presented. In Section \ref{s:chemmod}, we describe the main characteristics of the  chemical evolution model adopted in this work.
In Section \ref{results}, we present our Results, and finally in Section \ref{conc}, we draw our Conclusions.

\section{The \gaia DR3 sample in the solar vicinity}\label{s:gaia}
 In this Section, we provide all the information on the samples  of \gaia DR3 stars adopted in this study. In Section \ref{GSP}, we present the selection criteria for different chemical elements, in Section \ref{kordo} we briefly remind how the stellar ages proposed by \citet{Kordopatis2022} have been computed. Finally, in Section \ref{chemo_dyn} we summarise some of the dynamical and chemical properties of the selected \gaia DR3 stars considered here along with the distribution of associated \citet{Kordopatis2022} stellar ages.

\subsection{ Selection of solar vicinity stars and  {\it GSP-Spec} quality flags }
\label{GSP}
As mentioned in the Introduction, \citet{recioDR32022a} presented
 the largest homogeneous spectral analysis performed so far.
In this  work, we  are mainly interested  in the study of $\alpha$ chemical elements in the solar vicinity.
Taking  advantage of the large  number of stars in \gaia DR3, we are capable to limit our analysis
to a narrow region centered at the solar vicinity i.e. we consider stars with guiding radii $R_g$ $\in$ [8.1, 8.4] kpc (the motivation of this cut is  presented in the next paragraph). 
As in \citet{recioDR32022a}, we adopted the Sun’s Galactocentric position $(R,Z)_{\odot}=(8.249,0.0208)$ kpc \citep{gravity2021,bennett2019} 
and guiding radii have been computed using the rescaled version of the \citet{mcmillan2017} axisymmetric Galactic potential considering as inputs  the 
 geometric  distances by \citet{bailer2021}  based on high precision astrometric parameters from \gaia EDR3
\citep{brown2021} and the  additional  information provided by  \gaia DR3 for the radial velocities \citep{katz2022,vallenari2022}.

The selection of \gaia DR3 stars based on their guiding radii is motivated by the fact that in this way we  minimise the 'blurring', i.e. the orbit scattering due to interactions with disc  inhomogeneities and non-axisymetries (see \citealt{lynden1972,schoenrich2009MNRAS})
  which increase a star’s epicycle amplitude potentially without changing its angular momentum.   
Nevertheless, we are aware that  processes like the  'churning'
\citep{sellwood2002,schoenrich2009MNRAS} still could affect our stellar samples   selected by their guiding radii.
 In presence of  ‘churning’ the   angular momentum of the stars  changes and 
 moving them  from an almost circular orbit to another, thus erasing all memory of the birthplace of the star, based on its kinematics.
  For instance, in \citet{kordo_binney_2015} it was  shown that a significant number of super-solar metallicity stars have clear signatures of migrators into the Solar neighbourhood: those stars have circular orbits and kinematic ages that indicate they are at least few Gyr old.  

It is worth mentioning that the stellar sample selection based on a small range of  guiding radii values centered at the solar position also  helps to remove the effects of  abundance gradients along the Galactic disc \citep{kordopatis2020}.
Furthermore, we selected   only  giant stars with surface gravities log g < 3.5, g in cm s$^{-2}$. 
As clearly shown in the Kiel diagrams of  the Medium Quality  sample selected in different Galactic regions across the Milky Way as reported  in Fig. 6 of  \citet{recioDR32022b},  red giant stars  are observed in all the considered  vertical height bins above the plane, hence minimizing distance-dependent changes in the population being analysed.

  The first 13  parameter flags  of Table 2  of  \citet{recioDR32022a}  are shared by all the stars and have been set equal to 0, hence imposing the best  quality stellar atmospheric parameters. We refer the reader to \citet{recioDR32022a}  for all  details on the physical meaning of the flags. 
Concerning the single element quality flags for Si, Ca and Ti   we impose only the value  0 (best data) both for the flags 'UpLim' and 'Uncer'. Solely for the Mg, we also allow  Medium quality (0 and 1 values) for 'UpLim' (but still  0 for 'Uncer') because of the smaller number of stars available for this element (see \citealt{recioDR32022a}).

The polynomial coefficients as reported in Table 4 of \citet{recioDR32022a}   for the  log g versus [X/Fe] calibration (see their eq. 3) were applied considering  the suggested   validity domain in gravity.  Similarly, the  log g versus [M/H] calibration of eq. (2) in \citet{recioDR32022a}  (and relative coefficients reported in their Table 3) has been considered.

\subsection{Stellar ages}
 \label{kordo}
 
  \citet{Kordopatis2022} presented four different sets of ages and masses obtained through an isochrone fitting method.
  They used the  PAdova and TRieste Stellar Evolution Code (PARSEC) stellar tracks \citep{bressan2012}  up to the beginning of AGB phase and the  tracks up to the end of the AGB phase from  COLIBRI S37 (see  references in \citealt{Kordopatis2022}).
  In Section 2.2 of \citet{Kordopatis2022} the main methodology of the projection is reported in details. Here, we just recall that four following different combinations of spectroscopic parameters and broad-band photometric measurements (J,H,Ks from 2MASS and G from GDR3) were considered:
  
 \begin{itemize} 
 
 \item spec: projects only Teff, log g and [M/H].
  \item speck: projects Teff , log g, [M/H] and Ks.
\item specjhk: projects Teff, log g, [M/H], J, H and Ks.
\item specjhkg: projects Teff, log g, [M/H] J, H, Ks and G.
  \end{itemize}  
In the study by \citet{Kordopatis2022}, they also provided for the optimal combination of projection flavour as a function of the line of sight extinction in order to get the most reliable ages and masses.
   In this work, we compare our chemical evolution model predictions with only   \gaia DR3 stars with  associated ages characterised by relative errors smaller than 0.5 as suggested by \citet{Kordopatis2022}.

 In our study, it is important to be able to identify young stars as robustly as possible. \citet{Kordopatis2022},  
comparing the estimated ages of individual stars in open cluster members with the more robust  ages of \citet[][derived fitting the entire CMD of the cluster]{cantat2020},  found an over-estimation of 0.6 Gyr (inner dispersion of 1 Gyr) for cluster stars younger than 2 Gyr. This is a good starting point for our analysis, keeping in mind that some contamination might exist.

Based on this, we are confident that this not affects our conclusions,  since i)  we  included  a dispersion in the age estimates in our  model results (see Section \ref{3inf_details}) ii) we compare model prediction for young SSPs with also the Massive stars from \gaia DR3.

\section{The three-infall chemical evolution model for the  solar vicinity }\label{s:chemmod}
In this Section, we present the main assumptions  and characteristics of the chemical evolution model considered in this work for the Galactic region centered at the solar position.
In Section \ref{ref_2}, we briefly provide some details about the revised  two-infall model proposed  by \citet{spitoni2019,spitoni2020,spitoni2021} for the solar neighbourhood.
In Section \ref{3inf_sub}, we  postulate that an additional  third  accretion gas infall episode mimics the recent enhanced SF episodes found analysing \gaia data in the solar vicinity. In Section \ref{3inf_details}, we  provide all the details of the {\it three-infall} chemical evolution  model presented here.

\subsection{The revised two-infall  model  }\label{ref_2}

\citet{spitoni2019} and ES20 presented chemical evolution models designed to fit the observed chemical abundance ratios and asteroseismic ages of the APOKASC stars \citep{victor2018}. This sample contained about 1200 red giants from an annular region of  2 kpc wide centred on the Sun.  The stellar properties for this sample were determined by fitting the photometric, spectroscopic, and asteroseismic observables  \citep{silvaaguirre2017,aguirre2021}.  
In agreement with  the classical two-infall model by \citet{chiappini1997}, the first gas infall is characterised
by a short  accretion timescale  ($t_1<<t_2$), however it was highlighted the presence of a significant delay $t_{\rm max}$ between the two-infall accretion episodes  ($\sim$ 4 Gyr). \citet{spitoni2020} confirmed later this delay by using a 
 Bayesian framework based on Markov chain Monte Carlo methods.

  An  observational evidence supporting this scenario has been presented by \citet{nissen2020}  analysing the ESO/HARPS spectra of local solar twin stars. They found that the age-metallicity distribution shows the presence of  two diverse populations characterised  by a clear age separation. The authors suggested that these two sequences may be interpreted as evidence of two episodes of accretion of gas onto the Galactic disc with quenching of SF in between them.
More recently, \citet[][hereafter ES21]{spitoni2021} presented  a multi-zone two-infall chemical evolution model with  quantitatively inferred  free parameters by fitting the APOGEE DR16 \citep{Ahumada2019} abundance ratios at different Galactocentric distances.
In particular, the model computed at  8 kpc constrained by the [Mg/Fe] and [Fe/H] ratios of about 9200 stars located in the annular region enclosed between 6 and 10 kpc and vertical height |z| < 1 kpc, confirmed the previous findings  of a  significant delay between the two gas infall episodes.

\subsection{Beyond the two-infall model: including a recent episode of SF}\label{3inf_sub}
As it will be shown hereafter, the classical two-infall model fails of reproduce part of the new Gaia data.
The three-infall model for the Galactic  chemical evolution was originally introduced by \citet{micali2013} in order to study halo, thick and thin disc components, separately.  Here,  we  adopt 
a similar formalism but taking into account a recent episode with enhanced SF.

In their recent  work,  \citet{lara2020} presented  the  detailed star formation history  for both the thick and thin disc stars of the 2 kpc bubble around the Sun inferred from  \gaia DR2  colour–magnitude diagrams.
 Several and narrow episodes of enhanced star formation activity were revealed, where the last two occurred  approximately 1.9 and 1.0 Gyr ago. The authors proposed that the timing of these peaks of star formation coincides with the  Sagittarius dwarf spheroidal galaxy  pericentre passages  which  triggered  formation  of new stars in the Galactic disc.

\begin{figure}
\begin{centering}
\includegraphics[scale=0.58]{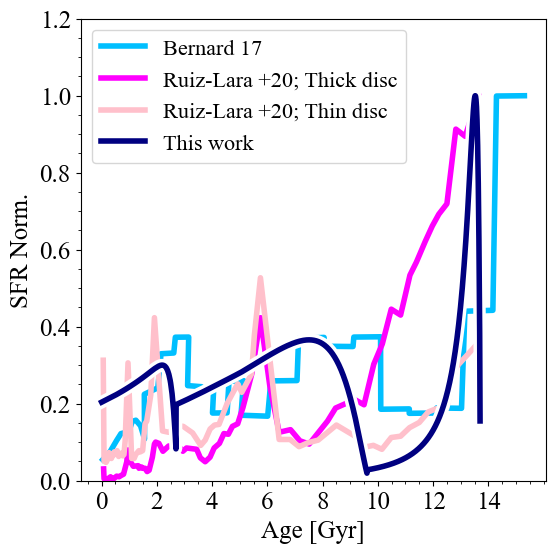}
\caption{The temporal evolution of the SFR normalised to the  maximum value predicted by our model is drawn with the dark blue line. The observed estimate   obtained  with \gaia DR1 data is reported by the light blue line  \citep{bernard2017}. The SF histories, based on  the second  \gaia release DR2   by  \citet{lara2020}, for the thick  and thin discs are reported with the magenta and pink lines, respectively.  }
\label{lara}
\end{centering}
\end{figure}

In Fig. \ref{lara}, we show the proposed star formation history  by \citet{lara2020} for both thick and thin disc sequences along with   \citet{bernard2017} results. In fact, we note that \citet{bernard2017} also highlighted the presence of  recent events with enhanced  star formation activity analysing    the first  \gaia data release.
This recent  enhancement of the  SF activity    can be easily mimicked in the chemical evolution models  by considering  a recent gas infall episode. In  Section \ref{dilution}, we will  discuss  that this assumption is also consistent with the abundance ratios observed in \gaia DR3 for massive stars and young objects.

\subsection{The three-infall model details }\label{3inf_details}
In this paper, we  extend the two-infall chemical evolution models presented in Section \ref{ref_2} in the light of the recent constraints imposed by the SFH  and  GDR3 chemical abundances.
 Considering as reference models the ones of Section \ref{ref_2},  we split  the low-$\alpha$ sequence  in  two distinct gas infall episodes  in order to mimic  the recent  enhanced SF activity of \citet{lara2020} and \citet{bernard2017}.
Here, in the framework of the  three-infall model, the functional form of the gas infall rate is:
\begin{eqnarray}\nonumber
\mathcal{I}_i(t)&=& \overbrace{\mathcal{X}_{1,i} \mathcal{A} \, e^{-t/ \tau_{1}}}^{\text{\textcolor{red}{{1st infall, high-$\alpha$}}}}+ 
 \overbrace{\theta(t-t_{{\rm max1}}) \, \mathcal{X}_{2,i} \, \mathcal{B} \, e^{-(t-t_{{\rm max1}})/ \tau_{2}}}^{\text{\textcolor{blue}{{2nd infall, low-$\alpha$ part I}}}}+\\
 && + \overbrace{\theta(t-t_{{\rm max2}}) \, \mathcal{X}_{3,i} \, \mathcal{C} \, e^{-(t-t_{{\rm max2}})/ \tau_{3}}}^{\text{\textcolor{blue}{{3rd infall, low-$\alpha$ part II}}}},
\label{infall}
 \end{eqnarray}
where $\tau_{1}$, $\tau_{2}$ and $\tau_{3}$  are the timescales of the three distinct gas infall episodes. 
The Heaviside step function is represented by $\theta$. 
  $\mathcal{X}_{1,i}$, $\mathcal{X}_{2,i}$ and  $\mathcal{X}_{3,i}$ are the abundance by mass unit of the element $i$ in the
infalling gas for the first, second and third gas infall, respectively.
The quantity $t_{{\rm max1}}$ is the time of the maximum infall rate on the  second accretion episode, i.e. it indicates the delay  between the two peaks of 
the thick disc and the  second infall (low-$\alpha$ part I) infall rates.
Similarly, the quantity $t_{{\rm max2}}$ is  the Galactic time  associated to the maximum infall rate of the  third accretion episode. 

Finally, the coefficients $\mathcal{A}$, $\mathcal{B}$ and $\mathcal{C}$ are obtained by imposing a
 fit to the observed current total surface mass density with  the following relations:

\begin{equation}
\mathcal{A} =\frac{\sigma_1}{\tau_{1} \left(1- e^{-t_{\rm max1}/\tau_{1}}\right)},
\label{S1}
\end{equation}

\begin{equation}
\mathcal{B} =\frac{\sigma_2}{\tau_{2} \left(1-
 e^{-(t_{\rm max2}-t_{\rm max1})/\tau_{2}} \right)},
\label{S2}
\end{equation}

\begin{equation}
\mathcal{C} =\frac{\sigma_3}{\tau_{3} \left(1-
 e^{-(t_G-t_{\rm max2})/\tau_{3}} \right)},
\label{S3}
\end{equation}
where $\sigma_1$ is  the present-day total surface mass density of the high-$\alpha$ sequence.  The $\sigma_2$ and $\sigma_3$ quantities  stand for the present-day total surface mass density of the two components of  the low-$\alpha$ sequence.
 Finally,  $t_G$ is the age of the Galaxy.
For the total present day surface density ($\sigma_1+\sigma_2 +\sigma_3$) in the solar neighbourhood, we assume  the value of 47.1 $\pm$ 3.4 M$_{\odot} \mbox{ pc}^{-2}$ as provided by \citet{mckee2015}
and used previously  by  \citet{spitoni2020,spitoni2021}.
The SFR is expressed as the \citet{kenni1998} law,
\begin{equation}
\psi(t)\propto \nu_{1,2,3} \cdot \sigma_{g}(t)^{k},
\label{k1}
\end{equation}
 where $\sigma_g$ is the gas surface
 density and $k = 1.5$ is the exponent. 
The quantity $\nu_{1,2,3}$ is the star formation efficiency (SFE) associated to different Galactic evolutionary phases. 
Motivated by the theory of star formation induced by spiral density
waves in Galactic discs \citep{wyse1989}, we consider a
variable SFE as a function of Galactic phase as already considered in the two-infall model
\citep{chiappini2001,grisoni2017,grisoni2019,grisoni2020,spitoni2020}.
We adopt the \citet{scalo1986}  initial stellar mass function (IMF), constant in
time and space.
Finally, we  adopt the photospheric values of \citet{grevesse2007} as our solar reference abundances, in order  to be consistent with the \gaia DR3 spectroscopic abundances.

Unlike other studies (e.g. \citealt{Nidever:2014fj}, for APOGEE data), in this paper we ignore the effect of Galactic winds on chemical evolution. 
 In fact, while studying the Galactic fountains originated by the explosions of Type II SNe in OB associations in the solar annulus, \citet{melioli2008, melioli2009} and \citet{spitoni2008, spitoni2009} found that the ejected metals fall back close to approximately the same Galactocentric region where they were ejected and consequently  do not modify significantly the chemical evolution of the disc as a whole. Furthermore,   the typical delay as computed by \citet{spitoni2009} of 0.1 Gyr (due to the orbit time of the clouds subject on the Galactic potential)   produces also a negligible effect on the chemical evolution of the Galaxy in the solar neighborhood.

In Section \ref{results}, we will show also   model results  considering - a posteriori - the  dispersion in the  abundance ratios and ages for the predicted simple stellar populations (SSPs, see also \citealt{spitoni2019}). 
We added, at each Galactic time, a random error to the ages and  ratios  [M/H], [X/M] 
of the SSPs
formed at Galactic evolutionary time $t$ as follows:

\begin{equation} 
\small
 \mbox{Age}_{new}(t)   = \mbox{Age} (t)  +  \delta_G(\mbox{Age});  \, \mbox{  } \delta_G( \mbox{Age})  \sim  \mathcal{N}(0, \sigma_{ \mbox{Age}})
\label{age_er}
\end{equation}
where $\delta_G$ is a perturbation which follows a Normal distribution $\mathcal{N}(0, \sigma_{Age})$ with the standard deviation fixed at the value of  $\sigma_{ \mbox{Age}}=30\% \mbox{ Age}$,
where  Age $(t)$ = (13.7- $t$) Gyr. Similarly,  for  [M/H] and [X/M] we have, respectively:

\begin{equation} 
\small
 \mbox{[M/H]}_{new}(t)   = \mbox{[M/H]} (t)  +  \delta_G(\mbox{[M/H]});  \, \mbox{  } \delta_G(\mbox{[M/H]}))  \sim  \mathcal{N}(0, \sigma_{ \mbox{[M/H]}}),
\label{mh_er}
\end{equation}

\begin{equation} 
\small
 \mbox{[X/Fe]}_{new}(t)   = \mbox{[X/Fe]} (t)  +  \delta_G(\mbox{[X/Fe]});  \, \mbox{  } \delta_G(\mbox{[X/Fe]}))  \sim  \mathcal{N}(0, \sigma_{ \mbox{[X/Fe]}}).
\label{xfe_er}
\end{equation}
In Eqs. (\ref{mh_er}) and (\ref{xfe_er}), we impose that  $\sigma_{ \mbox{[M/H]}} =\sigma_{ \mbox{[X/Fe]}} \equiv 0.05$ dex.
In the remainder of the article we will refer to this chemical evolution model
as our {\it synthetic model}.

 \begin{table*}
\begin{center}
\tiny
\caption{Summary of the main parameters of the best model presented in this study: the accretion timescales ($t_1$,  $t_2$, $t_3$),  time-delays  ($t_{{\rm max}1}$, $t_{{\rm max}2}$),
the present-day total surface mass density ratios ($\sigma_{2+3}$/ $\sigma_{1}$ and $\sigma_{2}$/ $\sigma_{3}$) and the SFEs ($\nu_1$,$\nu_2$, and $\nu_3$).  We also indicate the values that are identical to ones adopted in the previous two-infall chemical evolution models of ES20 and ES21.
}
\label{tab_3I}
\begin{tabular}{c|cccccccccc}
\hline
  \hline
\\
 &    $t_{1}$& $t_{2}$ &$t_{3}$&
$t_{\rm max1}$&
  $t_{\rm max2}$&
  $\sigma_{2+3}$/ $\sigma_{1}$&
 $\sigma_{2}$/ $\sigma_{3}$&
 $\nu_1$&$\nu_2$&$\nu_3$\\
 &[Gyr]& [Gyr]&[Gyr]&[Gyr]&[Gyr]& & &[Gyr$^{-1}$]&[Gyr$^{-1}$]&[Gyr$^{-1}$]\\
 \hline
 \\
  {\it Model  }& 0.103& 4.110 & 0.150&4.085& 11.000&3.472 &2.33 &2.0&1.3&0.5\\
 \\
&ES21 &ES21 &&ES21& &ES20& &ES21&ES20\\
\end{tabular}
\end{center}
\end{table*}

Although stellar migration has undeniably played an important tole in Galactic evolution such as the radial metallicity profiles  (e.g., \citealt{kordopatis2015}) and affecting the [$\alpha$/Fe]-age relation of thin disc stars \citep{vincenzo2020}, we decide in this work to ignore stellar migration effects in the Solar vicinity. Indeed, by means of  a self-consistent chemo-dynamical model for the Galactic disc evolution, \citet{Khoperskov2020} concluded that radial migration has a negligible effect on the [$\alpha$/Fe] versus [Fe/H] distribution over time, suggesting that the $\alpha$-dichotomy is strictly linked to different star formation regimes
over the Galaxy’s lifetime.  The only effect of migrators from the inner and outer disc Galactic regions would be to smear and smooth the disc bimodality. 

 Despite the presence of important signatures of the stellar migration,   \citet{vincenzo2020}  stressed  that two  distinct enriched    gas accretion episodes  with  primordial - or poorly  enriched - chemical composition occurred  at 0-2 and 5-7 Gyr ago,  defined  the  shape of the low-$\alpha$ sequence in the [$\alpha$/Fe] versus [Fe/H] plot. In their Fig. 13, it is worth noting that the abundances in stars younger than 8 Gyr in the solar neighbourhood perfectly trace the gas phase abundances in the same region.

Hence, we investigate a complementary scenario with respect to that proposed by \citet{sharma2020} and \citet{chen2022}, in which the dichotomy  visible in    [$\alpha$/Fe] versus [Fe/H] abundance  in the solar neighborhood   can  be entirely explainable by the stellar migration.

 The observed chemical impoverishment does not naturally appear in the sense of radial migration, that would favour a majority of stars coming from the chemically enriched inner galactic regions.

\subsection{Nucleosynthesis prescriptions}\label{nucleo}
In this article, we  assume the same stellar nucleosynthesis prescriptions as in ES20 and ES21, i.e. the ones of \citet{francois2004}. However, in the last part of the Results (Section \ref{romano_yields}),  we will also test  the effects  of the nucleosynthesis prescriptions suggested by \citet{romano2010}.

\subsubsection{  \citet{francois2004} yields collection  }\label{romano_yields}
 As in the models ES20 and ES21,    we adopt the same  nucleosynthesis prescriptions as proposed  by \citet{francois2004} for  Fe, Mg, Si, Ca and Ti.
  The authors artificially increased the Mg yields for massive stars from \citet{WW1995} to reproduce the solar Mg abundance.
 Mg yields from stars in the range 11-20 M$_ {\odot}$ have been increased by a factor of  7,
 whereas yields for stars with mass  $>20$ M$_ {\odot}$  are on average a factor $\sim$ 2 larger. 
 No modifications are required for the yields of  Fe and Ca, as computed for solar chemical composition. Concerning Si, only the yields of  very massive stars (M $>$ 40 M$_ {\odot}$) are increased by a factor of 2.  For the modification of Ti yields  we refer the reader to Fig. 7  of \citet{francois2004} where the  ratios between the revised yields  to the data  of \citet{WW1995} for massive stars are indicated. It is possible to note  that Ti yields have been substantially increased.
 Concerning  Type Ia SNe, in order to preserve the observed [Mg/Fe] and [Ti/Fe] versus [Fe/H] pattern, the yields of \citet{iwamoto1999} for Mg and Ti were increased by a factor of 5 and 2, respectively.
 The choice of such ad-hoc nucleosynthesis prescriptions is supported by the fact that stellar yields are still a relatively uncertain component of chemical evolution models \citep[e.g.][]{francois2004,romano2010,cote2017,prantzos2018}.
 This set of yields has been widely used in the literature \citep{cescutti2007,cescutti2022,spitoni2014,spitoni2015, spitoni2017,spitoni2D2018, mott2013, vincenzo2019,palla2022} and turned out to be able to reproduce the main features of the solar neighbourhood.

\subsubsection{ \citet{romano2010} yields collection}\label{romano_yields}
We also adopted  (see Section \ref{romano_yields})  an alternative set of stellar yields, coming from full stellar evolution and nucleosynthesis computations, i.e. those suggested by  \citet[][their model 15]{romano2010}
 and  lately  used in several Galactic archaeology studies   with chemical evolution models \citep[i.e.,][]{brusadin2013, micali2013, spitoni2016,spitoni2018}.
We decided to include this study in our analysis because, 
in contrast with \citet{francois2004},  \citet{romano2010} suggested a collection of yields without applying any modifications or tuning.

 For low-and intermediate-mass stars (0.8-8 M$_{\odot}$),
 the metallicity-dependent stellar yields of \citet{karakas2010} with thermal pulses  were included. 
For  the progenitors of either Type II SNe  or HNe
 (massive stars > 11-13 M$_{\odot}$ depending on the explosion energy) they used  the metallicity-dependent He, C, N and O stellar yields, as computed with the Geneva stellar evolutionary code, which takes into account the combined effect of mass-loss and rotation \citep{meynet2002, hirschi2005,hirschi2007, ekstrom2008}; for all the elements heavier than O, they assume the  stellar evolution calculations by \citet{koba2006}.

\subsection{Best fit-model parameters}\label{disc_results}
As already   explained in Section \ref{3inf_sub}, our main goal  is to extend the previous models presented  in ES20 and ES21 (for the solar vicinity) in the light of the new constraints given by the SF history  (from  \gaia DR1 and \gaia DR2 data) and new RVS chemical abundances ratios  provided by \gaia DR3. 
In Table \ref{tab_3I},  we summarise  the main parameters of the best model presented in this study.
Because the models ES20 and ES21 were already able  to successfully reproduce data from high resolutions survey such as  APOKASC  and APOGEE DR16, 
we tried to keep most of the model parameters similar to the values proposed by these studies.

For sake of clarity, in Table \ref{tab_3I} we indicate which  parameters have not been modified: the infall time-scales $t_1=0.103$ Gyr (high-$\alpha$), $t_2=4.110$ Gyr (low-$\alpha$, part I), the delay between the first infall and the second one $t_{max1}=4.085$ Gyr, the star formation efficiency of the high-$\alpha$ sequence $\nu_1=2$ Gyr$^{-1}$ are taken from ES21 model.
Defining  $\sigma_{2+3}/\sigma_1$ as the ratio between the low-$\alpha$   and high-$\alpha$ present day total surface mass densities, we assume $\sigma_{2+3}$/ $\sigma_{1}$=3.472 as prescribed by ES20. Concerning    the present-day ratio between  low-$\alpha$ and  high-$\alpha$ stars, the actual observed value is still uncertain, i.e.  we recall that \citet{fuhr2017} derived  in the solar vicinity a local mass density  ratio between the thin and thick disc stars of 5.26, which becomes as low as 1.73 after correction for the difference in the scale height. 
Finally,  the SFE of the low-$\alpha$ part I sequence is also taken from ES20 model ($\nu_2=1.3$ Gyr$^{-1}$).

As it will be discussed in Section \ref{results}, the RVS chemical abundance shows the presence of a population composed by massive stars  with a deficiency in metallicity [M/H], i.e. subsolar values  not explainable with standard chemical evolution or stellar migration processes (the same behavior is seen in young objects  with \citealt{Kordopatis2022} ages smaller than 1 Gyr). 
It is important to recall here that it has been found that this population traces the spiral arms.

In order to model correctly this population  we need for the third infall (low-$\alpha$, part II)  associated with lower SFE compared the first component of the low-$\alpha$ phase coupled with enough infall of gas able to produce sufficient dilution on short time-scales. In fact we fixed the SFE at the value of $\nu_3=0.5$ Gyr$^{-1}$, imposing that the mass of the part II is
$\sigma_3= 0.3  \cdot \sigma_{2+3}$
and  $t_3=0.15$ Gyr (see Table \ref{tab_3I} and the discussion in Section \ref{results} with the effects on the chemical evolution of this set of model parameters).

Finally, in order to better reproduce the \gaia DR3 abundance ratios, we impose for the two low-alpha components that the infalling gas 
 has a  mild chemical enrichment obtained from the model fixed at the value of  the high-$\alpha$ disc phase  corresponding to  [Fe/H]=-0.75 dex divided by a factor of 5.

\begin{figure}
\begin{centering}
\includegraphics[scale=0.25]{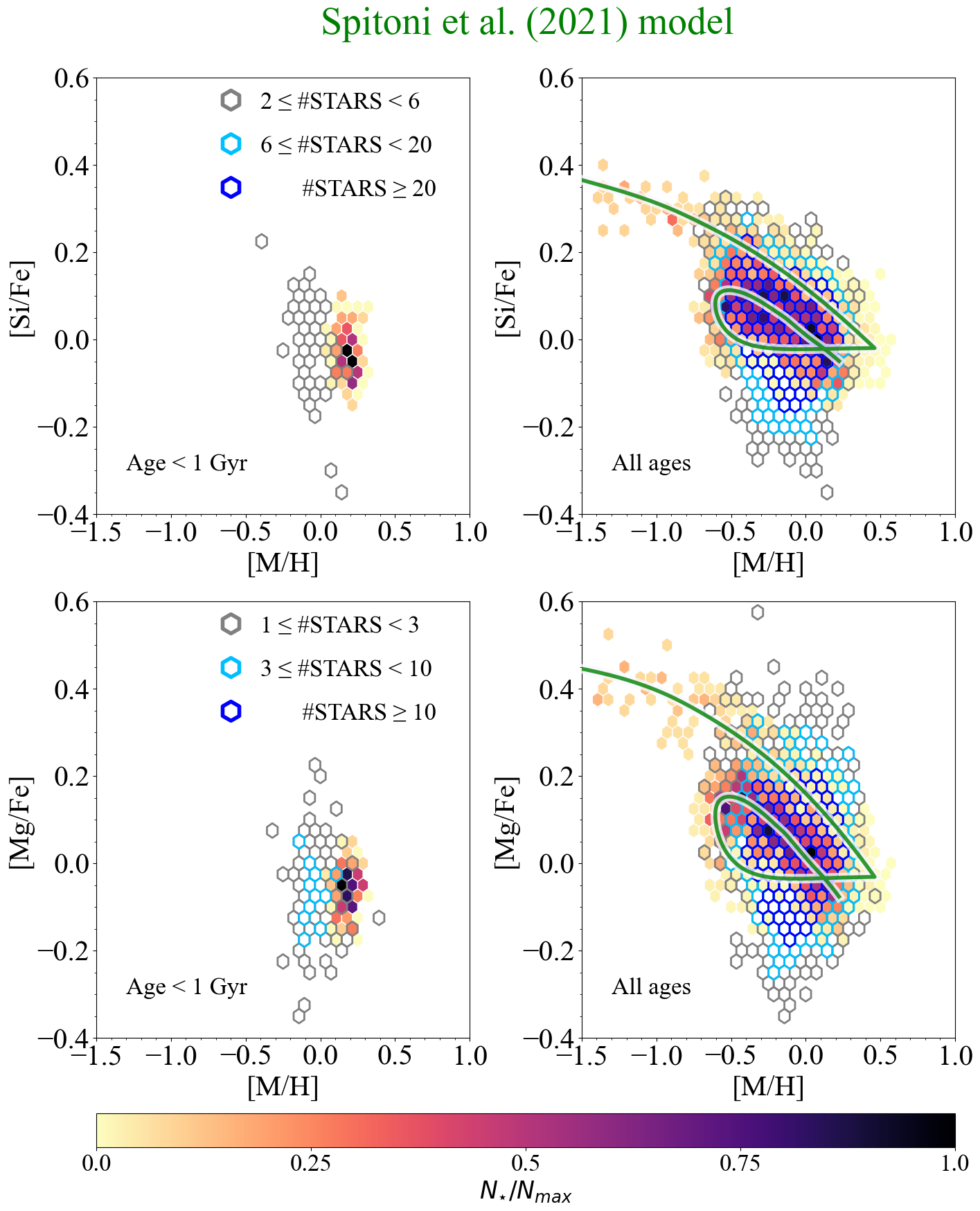}
\caption{Predictions  of the two-infall  model by ES21 including errors in the chemical abundance and age determinations (synthetic model) for the [Si/Fe]  (first row) and [Mg/Fe]  (second row) versus [M/H] relations. 
The color-coding stands for the normalised predicted total number of formed stars in the [Si/Fe] versus [M/H] (top panels) area covered by each hexagon.  $Gaia$ DR3 stars are indicated by  empty hexagons. For Si the grey ones enclose   between 2 and 6 stars, the light-blue between 6 and 20. Finally, blue  hexagons contain more than 20 stars.   In the lower panels, the numbers of \gaia DR3 stars contained in the hexagons for Mg is smaller  by a factor of 2  compared to  the Si plots (upper panels). 
In the left panels only stars with ages younger than 1 Gyr are represented, whereas in the right ones the whole ages are considered. We also depict the chemical evolution line predicted by  ES21 without considering any errors with the green lines.}
\label{es21}
\end{centering}
\end{figure}
\begin{figure}
\begin{centering}
\includegraphics[scale=0.27]{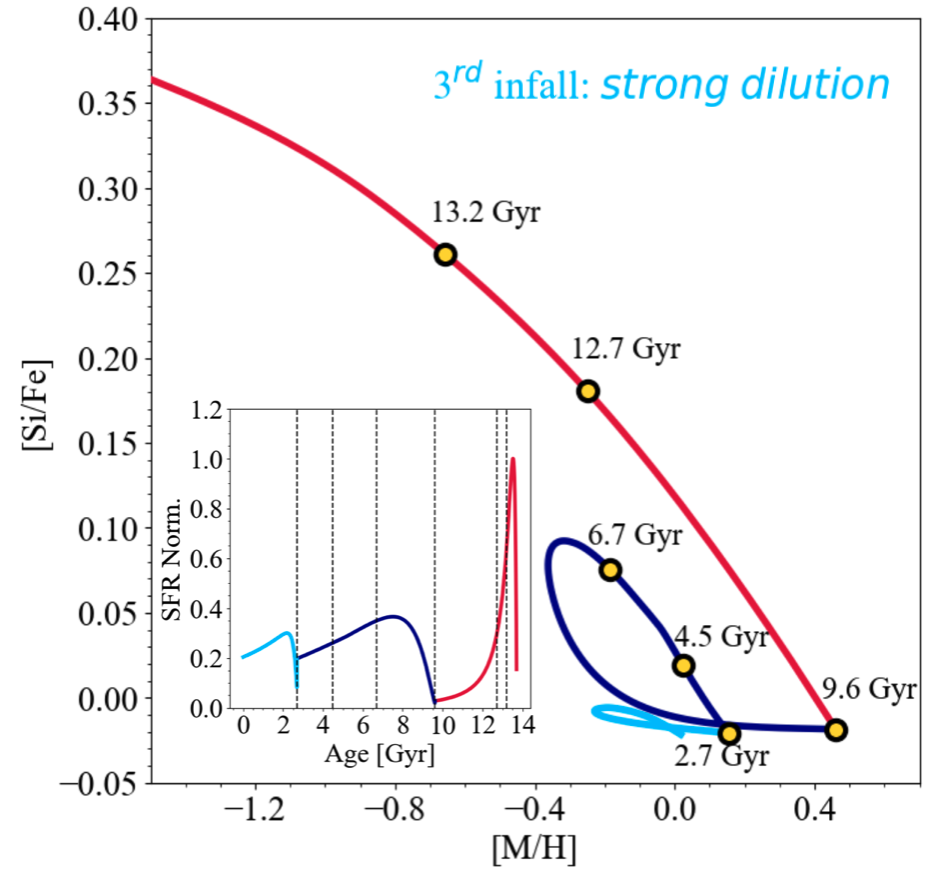}
\caption{The abundance ratios of [Si/Fe] as a function of [M/H] predicted by the three-infall chemical evolution model presented in Section \ref{s:chemmod} and with parameters listed in Table \ref{tab_3I}.
Different colors indicate the distinct infall episodes: red (high-$\alpha$), dark blue (low-$\alpha$, part I) and light blue (low-$\alpha$, part II). 
Filled yellow circles indicate the abundance ratios of the chemical evolution model at the given age. 
In the inset, the normalised star formation history is reported  with the different evolutionary phases colored as in main plot  [Si/Fe] versus [M/H]. With vertical dashed lines, we indicate the ages corresponding to   yellow points.}
\label{si_sfr}
\end{centering}
\end{figure}

\begin{figure}
\begin{centering}
\includegraphics[scale=0.27]{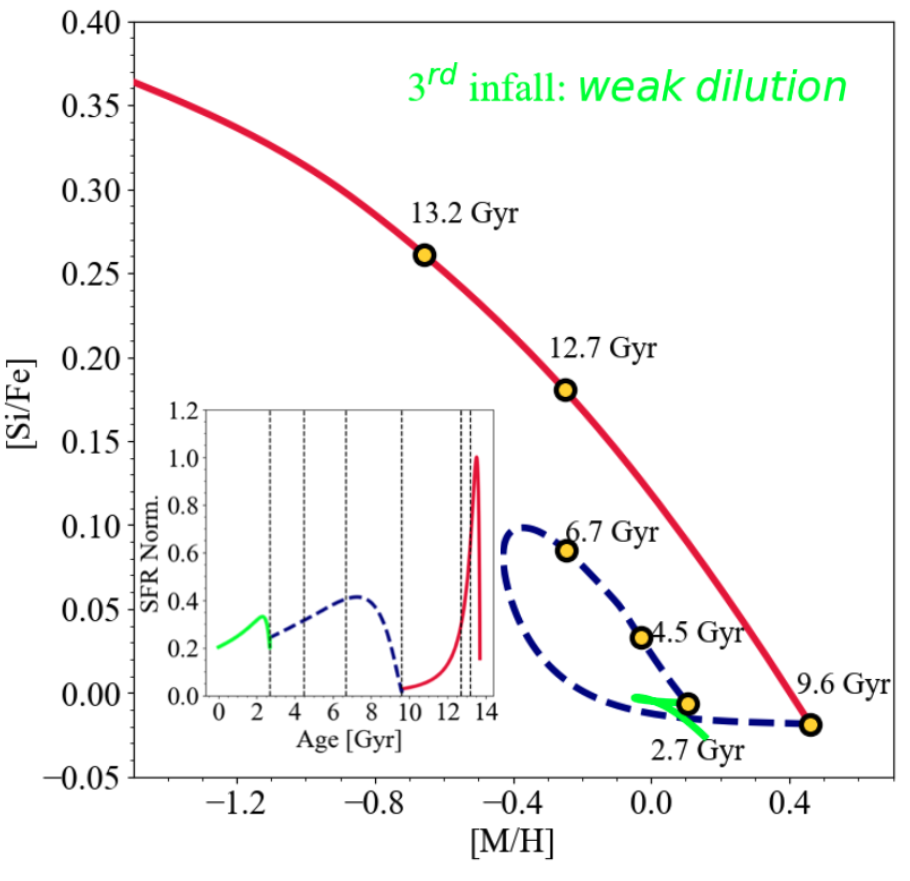}
\caption{  As in Fig. \ref{si_sfr}, but considering a smaller gas infall mass for the third infall - it accounts for 15\% of the total low-$\alpha$ sequence barionic mass - and a larger SFE ($\nu_3$=0.8 Gyr$^{-1}$). In the inset we also report the normalised
star formation history with the different evolutionary phases
colored as in main plot [Si/Fe] versus [M/H]. The  value of the  SFE  associated to  the low-$\alpha$ phase I  (indicated with the  blue dashed line ) is $\nu_2$=1 Gyr$^{-1}$.  Vertical dashed lines indicate the ages corresponding to   yellow points.    }
\label{si_sfr_nodilu}
\end{centering}
\end{figure}

  \begin{figure*}
\begin{centering}
\includegraphics[scale=0.5]{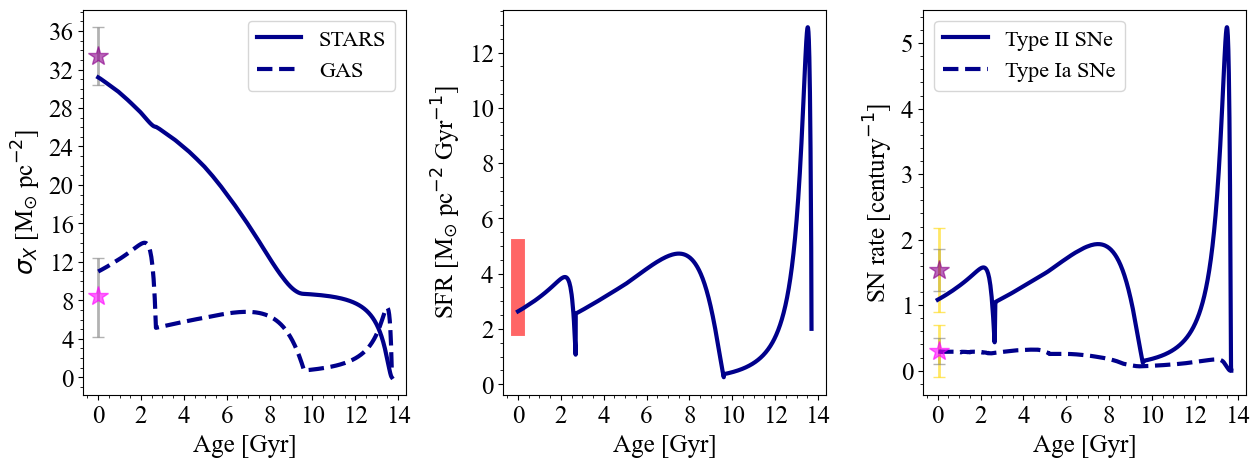}
\caption{ {\it Left panel}:  Surface mass density evolution of stars  ($\sigma_{\star}$, solid blue line) and gas ($\sigma_{g}$, dashed blue line) predicted by the  three-infall chemical evolution model presented in this study with model parameters reported in Table \ref{tab_3I}.  The purple star indicates the observed present-day $\sigma_{\star}$ value given by \citet{mckee2015}. The magenta star represents  the present-day  $\sigma_{g}$ value averaging between the \citet{dame1993} and \citet{Nakanishi2003,Nakanishi2006} data set as presented by \citet{palla2020} (here shown only at 8 kpc). 
{\it Middle panel}: The SFR time evolution predicted by the model is reported with the blue line. The red shaded area  indicates the measured range in the solar annulus suggested by \citet{prantzos2018}. {\it Right panel}: evolution of the Type Ia SN (dashed line) and Type II SN (solid line) rates predicted by model for  whole Galactic disc. The magenta star stands for the  observed Type Ia SN rate observed by \citet{cappellaro1997}, whereas the purple one stands for Type II SN rates observed by \citet{li2010}. 1$\sigma$ and 2$\sigma$ errors are reported with grey and yellow bars, respectively.  }
\label{SN}
\end{centering}
\end{figure*}

\section{Results} \label{results}

In this Section, we show the results of the new three-infall chemical evolution model introduced in Section \ref{s:chemmod} for $\alpha$-elements.  In Section \ref{res_twoinfall}, we compare \gaia DR3 data with previous two-infall models, for different chemical species.
In Section \ref{ref_3loop}, the main features of the proposed three-infall chemical evolution in the  [X/Fe] versus [M/H] (X=Mg, Si, Ca, Ti, $\alpha$)  plane are discussed.
In Section \ref{results_SFH}, we will present results of the temporal evolution of the SF and other Galactic disc observables. 
In Section \ref{age_bins}, we compare model prediction with \gaia DR3 data  for different age bins. In Section \ref{dilution}, we  show the dilution signature present in \gaia DR3 massive stars.

\subsection{The two-infall model and \gaia DR3 abundance ratios}\label{res_twoinfall}
In the right panels of Fig. \ref{es21},  we present the predictions of the reference two-infall ES21  model (green lines) compared with \gaia DR3 data.
As  illustrated in ES20 and ES21, the gas dilution originated by a strong second gas infall is a
key process to explain  APOKASC and APOGEE DR16  abundance ratios. In particular, the  second accretion  event of  pristine gas (or with a metal poor chemical composition)  decreases the metallicity of the stellar populations born immediately after  keeping  a roughly constant [X/Fe] ratio, where X is the $\alpha$-element.
When star formation resumes,  Type II SNe pollutes the ISM  producing a steep rise of the  [X/Fe] ratio, which subsequently decreases at higher metallicities due to bulk of  Fe injection    from Type Ia SNe \citep{matteucci2009, bonaparte2013, palla2021}. This sequence produces the characteristic  'loop'  feature, in the chemical evolution track of [X/Fe] versus [Fe/H]. 

In Fig. \ref{es21},  we also report the predictions of the two-infall ES21 'synthetic' model including errors for ages younger than 1 Gyr and the whole age range for  the [Si/Fe] and [Mg/Fe] versus [M/H]  relations. 
We have considered Si and Mg    because i) in ES20 with they approximate the global $\alpha$ ii) in ES21 the chemical evolution of Mg was studied.

Concerning the whole ages case (reported in the right panels), we can appreciate that the general trend of the data is well reproduced for the high-$\alpha$ and low-$\alpha$  phases. This agreement  is quite comforting,  because all of the parameters are the same as in ES20, ES21 for the first two gas infall (see Table \ref{tab_3I}).

However, it is clear  from  Fig. \ref{es21} that the ES21 model is not capable to reproduce the young population centered at under-solar [M/H] values predicting SSPs  with a too high metallicity.
The presence of a metal deficient young population in \gaia DR3 data  will be discussed in more details  in Section \ref{dilution}. 
 For the moment, we just mention that \citet{recioDR32022b}  pinpointed low [Mg/Fe], [Ca/Fe], and [Ti/Fe] ratios in the range -0.5 dex  < [M/H] < 0 dex for evolved massive stars near the disc plane.

This seems to indicate that the young stellar populations are chemically impoverished, and in the which can be  explained  with our three-infall model as we will see in Sections \ref{age_bins}  and \ref{dilution}.
In the proposed scenario the third infall can be the result of the dynamical interaction between the Milky Way and Sagittarius, as suggested by \citet{lara2020}.   As stated by \citet{recioDR32022b},  stellar migration  will not be able to explain this result either
because in such a scenario these stars should predominantly migrate from the outer disc but the decreasing stellar density with Galactocentric radius cannot allow them to dominate the star counts. 

\subsection{The three-infall model and the  dilution features}\label{ref_3loop}

In Fig. \ref{si_sfr}, we show the chemical enrichment history in the [Si/Fe] versus [M/H] plane for the  three-infall model without the inclusion of errors. We highlight   different phases associated to the distinct infall episodes with red (high-$\alpha$ sequence), blue (low-$\alpha$ part I) and light-blue (low-$\alpha$ part II) lines to visualise better the main trends.
Soon after the injection in the Galactic system of  the infalling gas associated with the first component of the low-$\alpha$ sequence (low-$\alpha$ part I in eq. \ref{infall}), the  dilution phase begins.
In particular, as mentioned above for the ES20 and ES21 model,  the  second accretion  event of metal poor gas decreases the metallicity of the stellar populations born immediately after  keeping  a roughly constant [Si/Fe] ratio.
Once the SF resumes, the predicted Galactic chemical evolution follows the characteristic  'loop' feature in the [Si/Fe] versus [Fe/H] plane.
 In \citet{johnsom2020} and \citet{lian2020b, lian2020} the effects of a recent gas rich merger or violent dynamical disturbance  on the chemical evolution have been discussed producing features similar to the ones described above in the chemical space. 

\begin{figure*}
\begin{centering}
\includegraphics[scale=0.3]{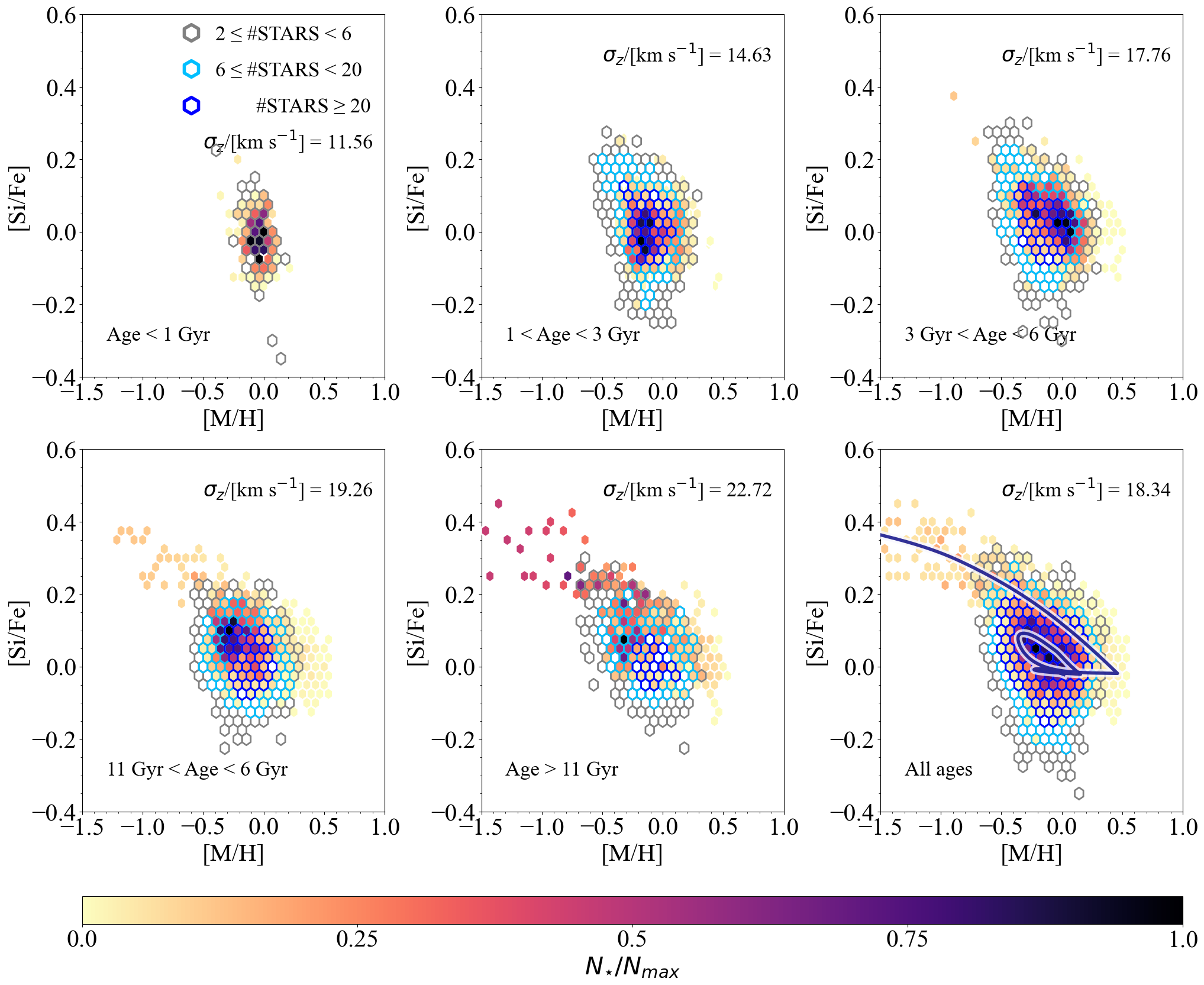}
\caption{The abundance ratios of [Si/Fe] as a function of [M/H] predicted by our three-infall 'synthetic' chemical evolution model for different age bins  taking into account observational errors (see  Eqs. \ref{age_er},  \ref{mh_er} and \ref{xfe_er})  are indicated with fully colored-coded hexagons.
The color-coding stands for the normalised predicted total number of formed stars in the [Si/Fe] versus [M/H] area covered by each hexagon. 
$Gaia$ DR3 stars are indicated by  empty hexagons: the grey ones enclose   between 2 and 6 stars, the light-blue between 6 and 20. Finally, blue  hexagons contain more than 20 stars. 
In the third panel of the second row  we show model prediction for the all ages and we also indicate in blue the model results without the errors. In each panel we report as well  the median vertical dispersion velocity values computed at  different age bins.
}
\label{si}
\end{centering}
\end{figure*}

In Fig. \ref{si_sfr},   the effects  of the third accretion event  (with  model parameters reported in Table \ref{tab_3I}) on the chemical enrichment of the Galactic disc can appreciated.   It begins after 11 Gyr of evolution (Galactic age of 2.7 Gyr) and  produces  a much smaller 'loop' feature compared to the second infall.  
As clearly indicated by the light-blue line in Fig. \ref{si_sfr},  the principal effects of the  third gas infall  are: i) keeping the  present-day metallicity [M/H]  at value smaller compared the previous chemical evolution model proposed by \citet{spitoni2019,spitoni2020, spitoni2021} ii) maintaining low values of [Si/Fe]. In the first upper panel of Fig. \ref{si},  we can see that this third infall allows our 'synthetic' model to well reproduce  even the observed stars younger than 1 Gyr.

We have mentioned in Section \ref{disc_results}   that  the total barionic mass in the low-alpha sequence (i.e. the $\sigma_{2+3}$ quantity in Table \ref{tab_3I}) is distributed as follows: 70\%  in the low-$\alpha$ part I and 30\% low-$\alpha$ part II phases. Hence, the smaller mass associated to the third infall combined with the  low SFE ($v_3= 0.5$ Gyr$^{-1}$) are the main responsible for the  small loop in [Si/Fe] versus [M/H]  relation. 
However, the dilution required to reproduce the young population at subsolar metallicity can be in place imposing a short  time-scale $t_3$ for the gas accretion. 
This particular combination of parameters allows us to predict the young subsolar metallicity population as we will show in Section \ref{age_bins} and \ref{dilution}.

 For sake of completeness, in Fig. \ref{si_sfr_nodilu} we  show a  chemical evolution model 
constrained by a similar SF history as the one reported in the inset panel of Fig. \ref{si_sfr}  that presents a different chemical enrichment in the [Si/Fe] versus [M/H] space.
We imposed a smaller gas infall mass for the third infall -it accounts in this case for the 15\% of the total low-$\alpha$ sequence baryonic mass- and a larger SFE,  $\nu_3$=0.8 Gyr$^{-1}$.

\begin{figure*}
\begin{centering}
\includegraphics[scale=0.3]{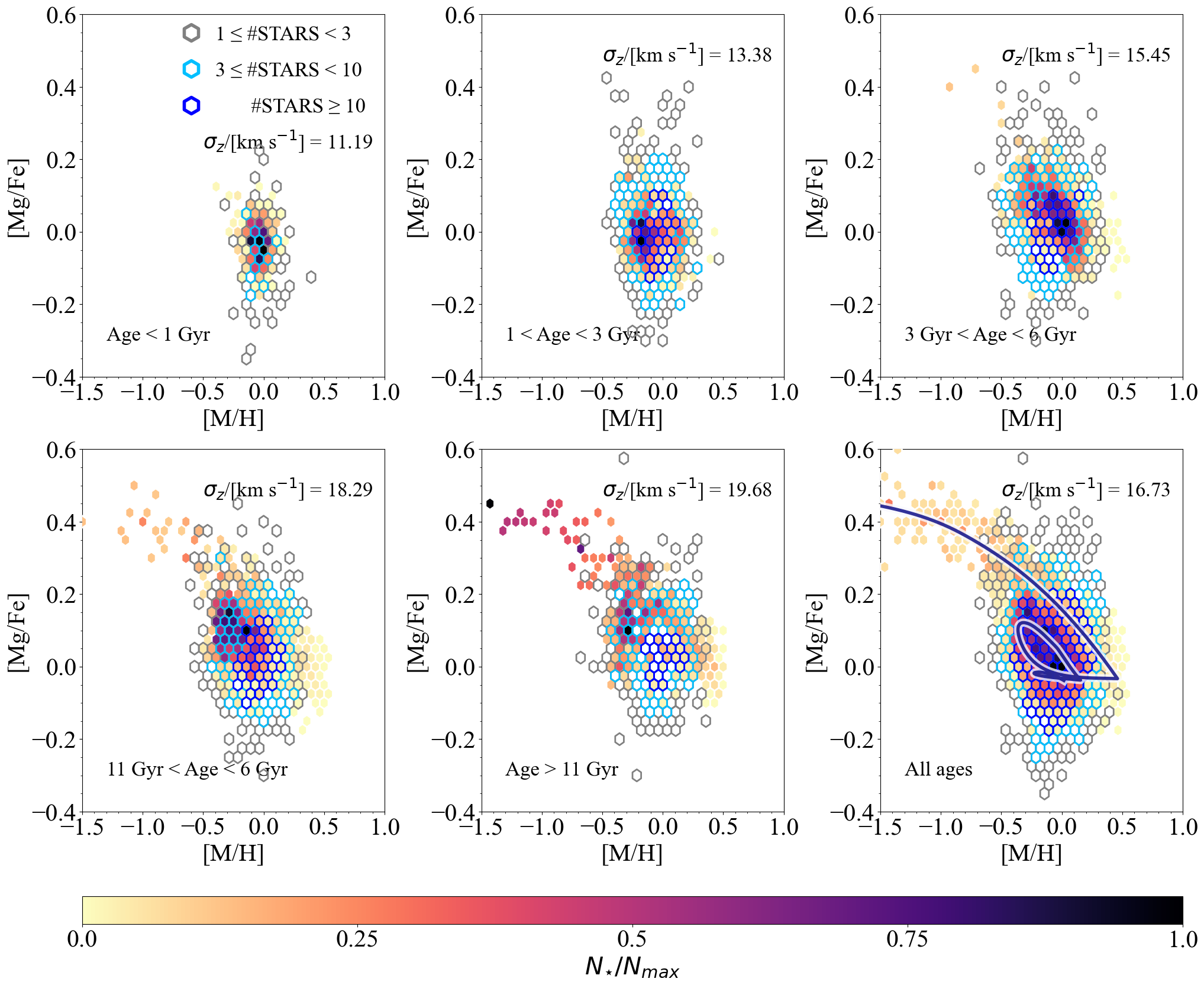}
\caption{ As in Fig. \ref{si} but for Mg. In this case, the  grey empty  hexagons contain a number of \gaia DR3 stars  between 1 and 3, the light-blue  ones between 3 and 10, and blue ones indicate regions with a number of stars larger than 10 stars. }
\label{mg}
\end{centering}
\end{figure*}

\begin{figure*}
\begin{centering}
\includegraphics[scale=0.3]{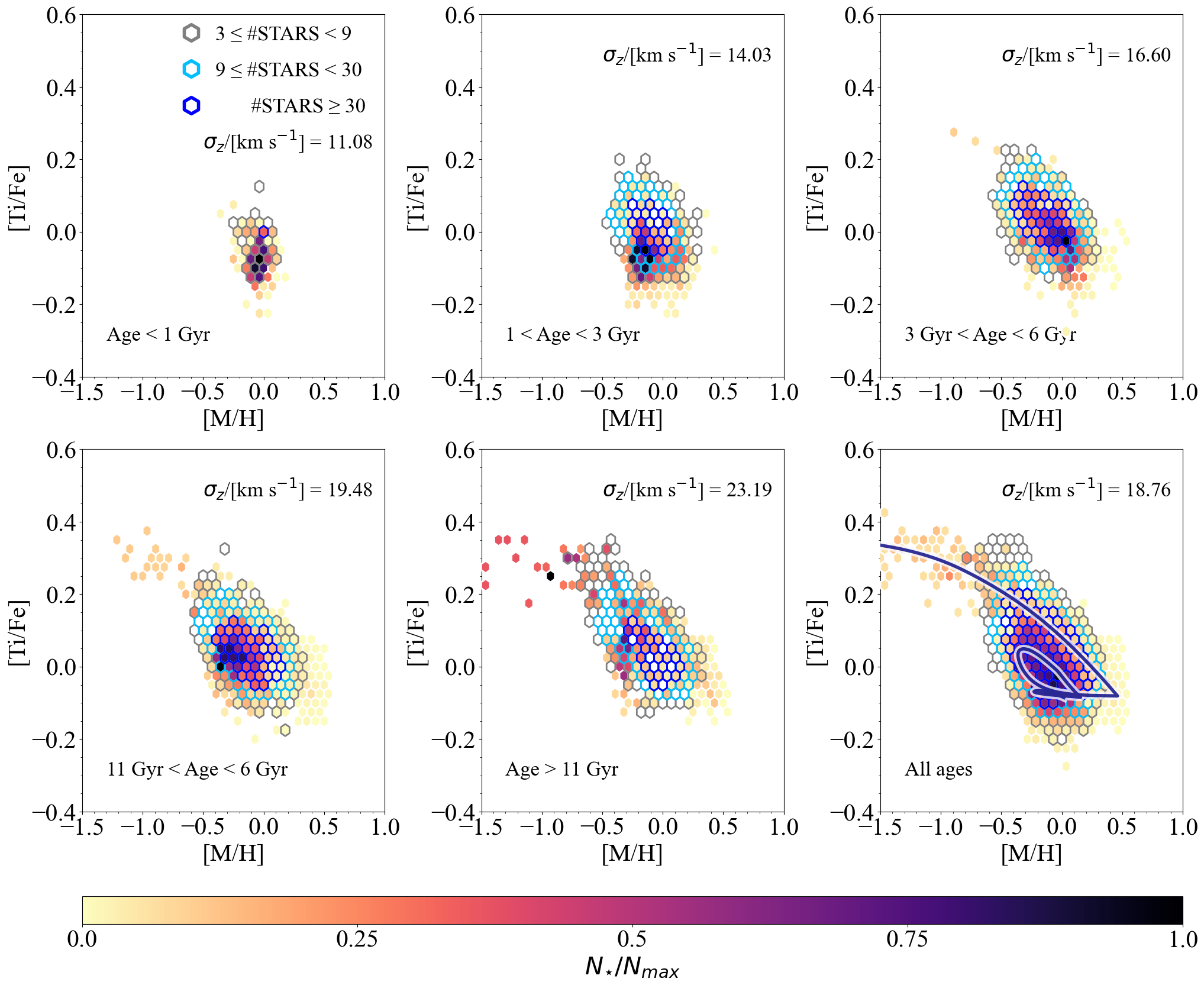}
\caption{ As in Fig. \ref{si} but for  Ti.  Here, the  grey empty hexagons   contain a number of \gaia-DR3 stars  between 3 and 9, the light-blue ones between 9 and 30, and blue  ones larger than 30.}
\label{ti} 
\end{centering}
\end{figure*}

In order to keep the value  of the ratio $\sigma_{2+3}/\sigma_{1}$  suggested by ES20 as  indicated in Table \ref{tab_3I}  imposing a lower mass associated to the third infall,  a larger mass for the low-$\alpha$ phase I) component is required. Hence,  we were able to reproduce the  SF history  of Fig. \ref{si_sfr},  assuming    a smaller   SFE value of the phase  low-$\alpha$ phase I:   $\nu_2$=1 Gyr$^{-1}$.  In  Fig. \ref{si_sfr_nodilu}, we note that a weaker chemical  dilution is present in the [Si/Fe] versus [M/H] space compared to the model of Fig. \ref{si_sfr} during the third infall phase, 
showing a chemical evolution track very similar to the ES20 and ES21 models.   It is  worth mentioning that both models presented in Figs.  \ref{si_sfr} and  \ref{si_sfr_nodilu}  are also able to reproduce the data studied in the earlier works of ES20 and ES21. The proposed three-infall model with strong dilution has been used to interpret the [M/H] versus [Ce/Fe] \gaia DR3 abundance ratios in \citet{contursi2022}.

Future \gaia data releases, as well as other surveys, might revise the properties and significance of the young population of metal-impoverished stars found by \citet{recioDR32022b}.  However, {\it the three-infall model presented here, constrained by the recent local SFH derived from Gaia data, would still remain a viable scenario for the local disc formation.} It could be necessary, though, to assume a higher/smaller amount of accreted gas. In this respect, it has to be mentioned also that the identification of the young, metal-poor component could suffer from biases related to the adoption of the standard spectroscopic LTE analysis for those stars  (see discussions in  \citealt{magrini2022}).
Furthermore, the parameterization of young stars from high-resolution spectra (R$\sim $100,000) could be also affected by the  combination of intrinsic factors  such as activity, fast rotation, magnetic fields  \citep{zhang2021,spina2020,spina2022}.
 However,  \citet{recioDR32022b} pointed out that  the above-mentioned processes should be important in young stars with age  younger than 200 Myr, not affecting the \gaia  RVS parameterization, which is performed at medium resolution (R$\sim$11,500).

\subsection{The Galactic star formation history and other disc observables}\label{results_SFH}

In Fig. \ref{lara}, we already compared the  predicted SF history  by the model with parameters given in  Table \ref{tab_3I}  with  the ones of   by \citet{lara2020} and \citet{bernard2017}. We see that our three-infall model approximately traces the main feature of the observed SF history by \gaia, and that the recent narrow SF peaks  can be well mimicked by a low-$\alpha$ sequence formed by 2 independent episodes of gas infall. Concerning the high-$\alpha$   sequence, the predicted  SF is in very good   agreement with the  thick disc part of \citet{bernard2017}. On the other hand, in \citet{lara2020} we  notice that  the thick disc has a more extended evolution in time.

  We want to recall here, that also the Milky Way-like galaxy simulations in the cosmological framework   presented by \citet{vincenzo2020}  show important signatures of two  gas accretions  of metal-poor gas occurred   about 0-2 and 5-7 Gyr ago - {\it in very good agreement with  the scenario  we  proposed  with  three-infall model} -  responsible for the shape of the low-$\alpha$ sequence in the  [$\alpha$/Fe]–[Fe/H] diagram  for the  neighbourhood.

As mentioned in Section \ref{disc_results}, in this study we want to retain, especially for the high-$\alpha$ sequence, the model parameters of previous chemical evolution models already capable to reproduce appropriately abundance ratios of APOGEE and APOKASC data.

As anticipated in Section \ref{3inf_details}, we impose that  the total (sum of high- and low $\alpha$ sequence contributions) surface mass density in the local disc is the one suggested by   \citet{mckee2015} i.e.  47.1 $\pm$ 3.4 M$_{\odot} \mbox{ pc}^{-2}$. In the same article, the authors also provided the values  for  the present-day total local surface density of stars,  $\sigma_{\star}(t_g)=33.4$ $\pm$ 3 M$_{\odot} \mbox{ pc}^{-2}$. 
In the left panel of  Fig. \ref{SN}, we draw the temporal evolution of the surface density of stars predicted by our three-infall chemical evolution model. With this model we have $\sigma_{\star}(t_g)=31.2$ M$_{\odot} \mbox{ pc}^{-2}$, in agreement with the above mentioned value  proposed by \citet{mckee2015}.
In the same panel,  we show the predicted temporal evolution of the surface gas density. In particular, the computed present-day value of   $\sigma_g(t_g)$=11.0 M$_{\odot} \mbox{ pc}^{-2}$ is   higher than  the value computed by  \citet{palla2020} of 8.4$^{+4.0}_{-4.2}$ M$_{\odot} \mbox{ pc}^{-2}$ (averaging between the \citet{dame1993} and \citealt{Nakanishi2003,Nakanishi2006}  data sets) but still within 1$\sigma$ uncertainty.

In the  middle panel of Fig. \ref{SN}, we present the time evolution of the SFR in our model, which predicts a present day  value of 2.63 M$_{\odot}$ pc$^{-2}$ Gyr$^{-1}$, hence in excellent agreement with the range usually assumed as constraint in chemical evolution models in the solar vicinity of 2-5 M$_{\odot}$ pc$^{-2}$ Gyr$^{-1}$  \citep{matteucci2012,prantzos2018}.

In the right panel of Fig. \ref{SN}, we report  the time evolution of the Type Ia SN and Type II SN rates. The present-day Type II SN rate in the whole Galactic disc predicted by our three-infall model is 1.08 /[100 yr], a smaller value (but within 2$\sigma$ error) than the observations of \citet{li2010} which yield a value of 1.54 $\pm$0.32 /[100 yr]. 
The predicted present-day Type Ia SN rate in the whole Galactic disc
is 0.29 /[100 yr],  in very good agreement with the value provided by \citet{cappellaro1997} of 0.30$\pm$0.20  /[100 yr]. 
 Moreover,  the  computed present-day infalling gas rate   is 0.71 M$\odot$ pc$^{-2}$ Gyr$^{-1}$,  consistent with the  range 0.3 -1.5 M$\odot$ pc$^{-2}$ Gyr$^{-1}$ suggested in Chapter 5.3.1 of  \citet{matteucci2012} for the solar vicinity.

\begin{figure*}
\begin{centering}
\includegraphics[scale=0.3]{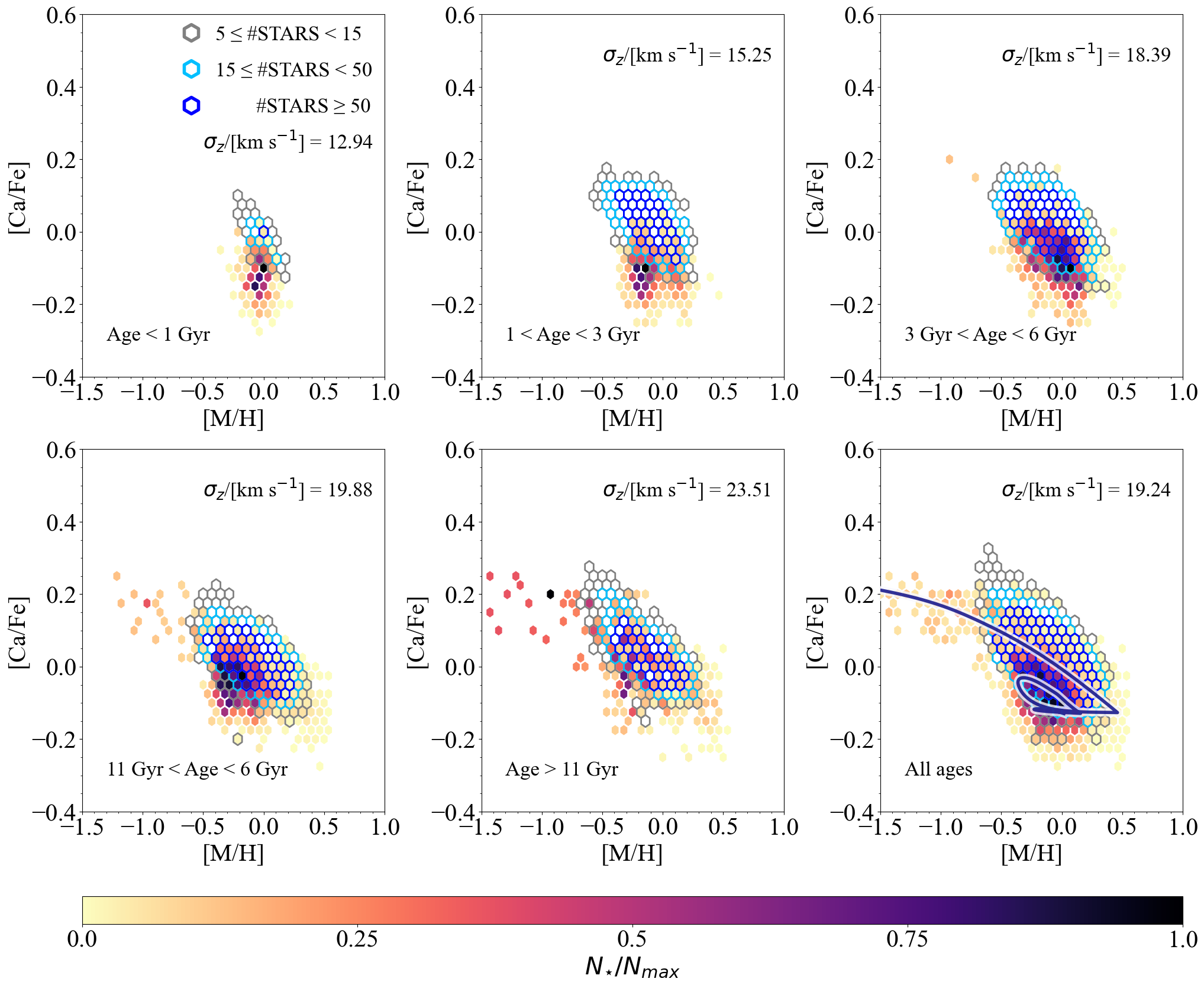}
 \caption{ As in Fig. \ref{si} but for Ca. In this case, the  grey empty  hexagons enclose a number of \gaia DR3 stars  between 5 and 15, the light-blue ones between 15 and 50, and blue  ones larger than 50.}
\label{ca} 
\end{centering}
\end{figure*}

 \begin{figure*}
\begin{centering}
\includegraphics[scale=0.23]{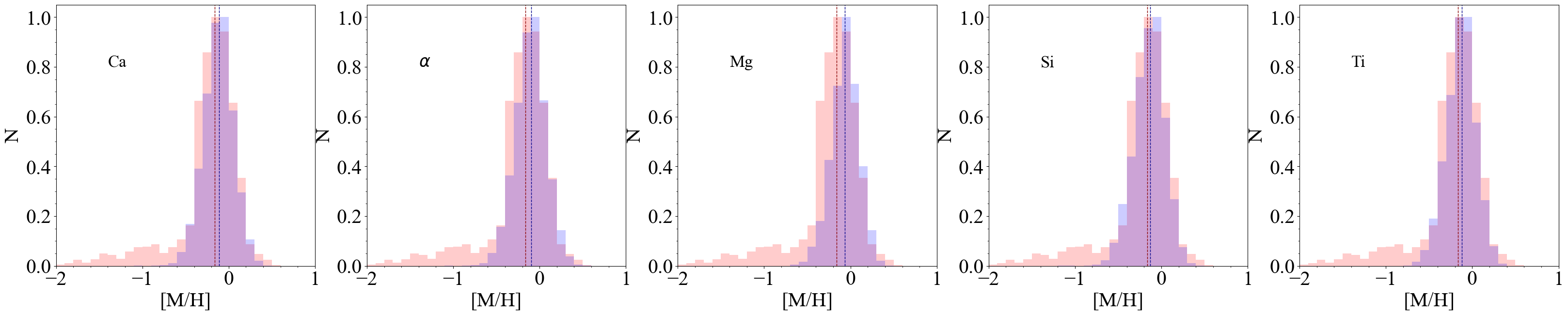}
\caption{
Metallicity [M/H] distributions predicted by the 'synthetic'  model (light-red histograms) compared with the ones obtained considering the different Gaia DR3 stellar sub-samples  for Ca, $\alpha$, Mg, Si and Ti, respectively (blue histograms). In each plot the associated median values are also indicated with the vertical dashed lines.}
\label{MDF}
\end{centering}
\end{figure*}

\subsection{Three-infall model predictions  in different age bins}\label{age_bins}

In this Section, we present the predicted abundance ratios  by the three-infall chemical evolution  and the associated 'synthetic'  models.
In Fig. \ref{si}, we report the [Si/Fe] versus [M/H] abundance ratios computed in six different age bins. 
We recall that in GSP-Spec   [M/H] values follow the [Fe/H] abundance with a tight correlation, in the following plots we compare [Fe/H] model predictions with [M/H] \gaia DR3 abundance ratios.

We can appreciate that our three-infall 'synthetic model' is able to catch the main features of \gaia DR3 data at different ages. In particular, we are able to well reproduce the youngest stellar population with ages < 1 Gyr.

In Fig. \ref{si}, we also indicate in each panel  
the median vertical velocity dispersion $\sigma_Z$ of the \gaia stars in each age bin. As expected the older age bins are characterised by larger   $\sigma_Z$  values.

In the lower right panel, we consider the whole age range and  beside the 'synthetic model'  it also drawn the  best-fit three-infall model of Fig. \ref{si_sfr} with  the blue line.  As already discussed in Section \ref{ref_3loop}, the chemical evolution track is very similar to the one previously proposed by \citet{spitoni2019}, ES20 and ES21: once the  high-$\alpha$ sequence ended (after $\sim$ 4 Gyr of evolution), 
the Galactic disc has achieved the highest value of metallicity [M/H].

In Fig. \ref{mg}, we show the model results for the [Mg/Fe] versus [M/H], and also in this case we find a very good agreement between our 'synthetic' model  computed at different age bin and \gaia DR3 data,  similarly to the results presented above for Si. In Fig. \ref{es21}, we also report ES21 model predictions for Mg compared with \gaia DR3 data. As for the Si, model ES21 (designed to reproduce APOGEE DR16 data in the solar vicinity) is capable to reproduce nicely  \gaia DR3 if we consider the whole  stellar ages  (right lower panel of Fig. \ref{es21}), but it cannot predict the \gaia DR3 stars younger than 1 Gyr  impoverished in metals [M/H]  (left lower panel).

 As shown in Fig. \ref{ti}, we also have   a good agreement between models and data for Ti, indeed the 'synthetic' model traces very well the main features of [Ti/Fe] versus [Fe/H] in all the considered age bins.
In Fig. \ref{ca}  we draw model results for [Ca/Fe] versus [M/H].  We can notice that the model  slightly   underestimates  [Ca/Fe] and  [$\alpha$/Fe]  \gaia DR3 abundance ratios ($\sim 0.05$ dex).
In Fig. 4 of  \citet{francois2004} it is possible to note that even for the corrected yield the predicted [Ca/Fe] versus [Fe/H] abundance ratios seems to be
below the collection of stellar data consistently with our results shown in Figs. \ref{ca}.

 In Fig. \ref{MDF}, we report the metallicity distribution function (MDF) predicted  by our 'synthetic' model compared with  ones computed for  different stellar samples selected for  different elements. We can appreciate that our 'synthetic' model MDF is in very good  agreement with  Ca, $\alpha$, Si and Ti stellar samples. Only for Mg, the \gaia DR3 stellar distribution  shows much less stars at low metallicity. 
In the \gaia RVS range the unique Mg line is very weak so the abundances have to be treated carefully.
For this reason we have  availability of a limited number of  stars with high quality values for the  flags is  the main reason of this mild discrepancy.

\subsection{Recent dilution signature in \gaia DR3 massive stars}\label{dilution}
As already pointed out by \citet{recioDR32022b}, evolved massive stars in the same parameter range as the Massive population presented in their Section 3,  dominate the observed populations visible for X= Mg, Ca, Ti  abundances in the range [M/H] $\in$ [-0.5, 0] dex, and at low abundance [X/Fe] ratio values near the disc plane at all Galactic radii.

In the previous Section, we have seen that this population of young stars seems to be impoverished in the Solar cylinder, as has been found by \citet{recioDR32022a}  considering a larger number of element abundances. It can be noted that silicon, calcium and titanium, which have several lines in the RVS wavelength range, have been measured in a large sample of stars in the solar cylinder: from more than 80,000 for [Ti/Fe] to more than 140,000 for [Si/Fe].

Moreover, in Appendix D of \citet{recioDR32022b},  the robustness of the Massive stars chemical impoverishment estimate and its dependence on the different calibrations have been also verified.

 \begin{figure*}
\begin{centering}
\includegraphics[scale=0.28]{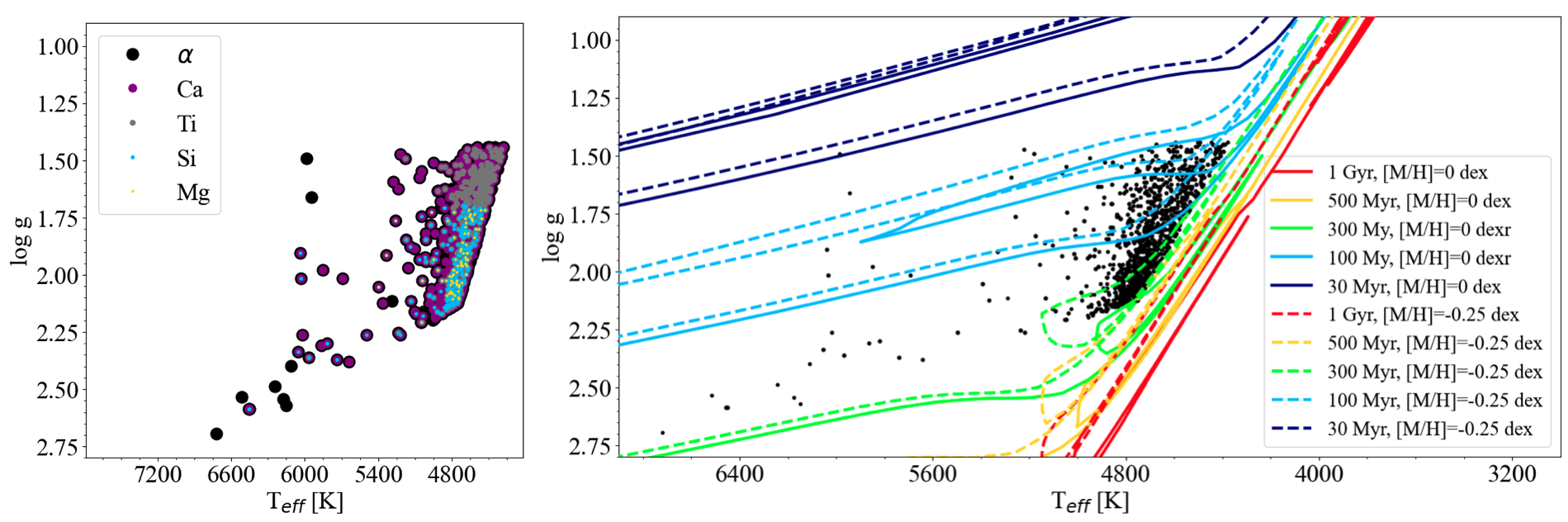}
\caption{ {\it Left panel:} Kiel diagram  for the massive \gaia DR3 stars  sub-samples selected as indicated  in Section \ref{dilution}  for $\alpha$, Ca,  Mg, Si and Ti, respectively. 
Different sizes of the point have  the only scope to visualise better the stars in common among  different sub-samples.  {\it Right panel:} PARSEC isochrones for    metallicites [M/H]=0 dex and [M/H]=-0.25 dex, compared to the massive \gaia DR3 stars sub-samples (black points). }
\label{kiel_massive}
\end{centering}
\end{figure*}

\begin{figure*}
\begin{centering}
\includegraphics[scale=0.25]{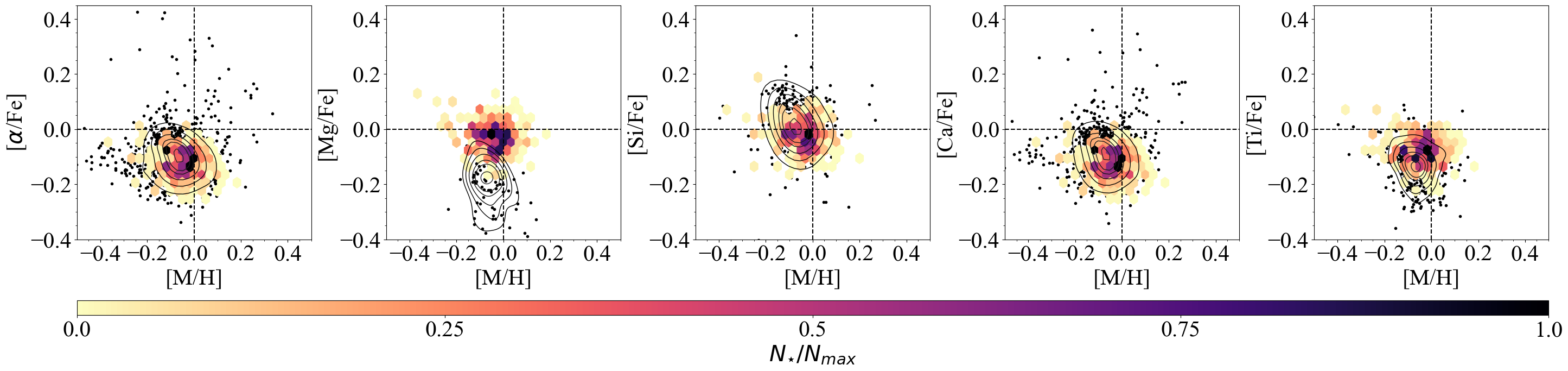}
\caption{   Comparison between abundance ratios of  [$\alpha$/Fe],  [Mg/Fe], [Si/Fe], [Ca/Fe],  [Ti/Fe]   as a function of [M/H]   predicted by the three-infall model for  stars younger than 1 Gyr  compared and \gaia DR 3 massive stars.
Black points indicate observed massive  stars  selected as in \citet[][see text for details]{recioDR32022b} using   the calibrations proposed by \citet{recioDR32022a}. 
  The contour
lines enclose fractions of  0.75, 0.60, 0.45, 0.30,0.20 and 0.05 of the total number of observed stars.  The colored-coded hexagons indicate model predictions highlighting the total number of stars formed by the fiducial three-infall model in the different region of the abundance ratio relation. }
\label{massive}
\end{centering}
\end{figure*}

 \begin{figure*}
\begin{centering}
\includegraphics[scale=0.4]{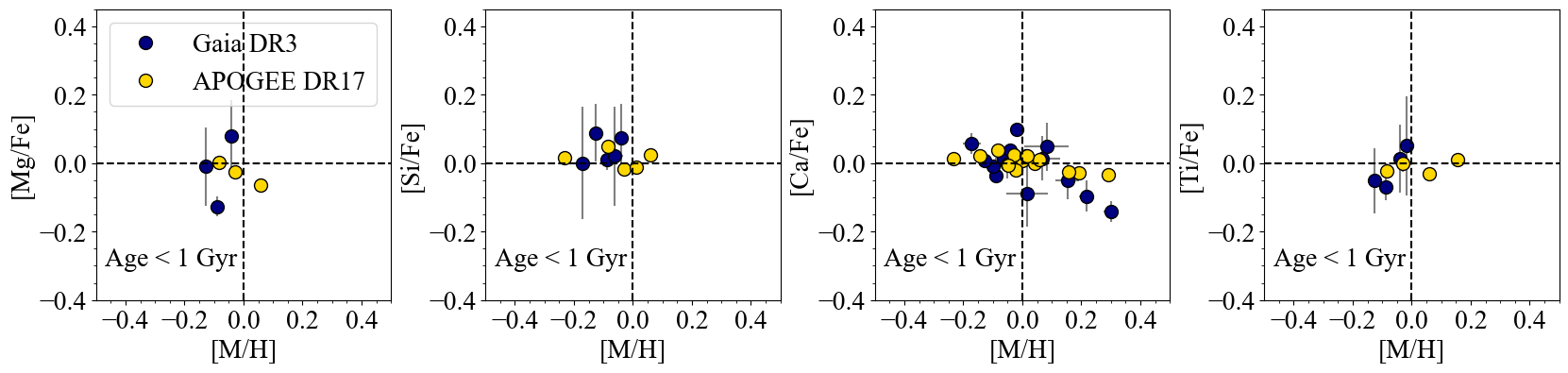}
\includegraphics[scale=0.4]{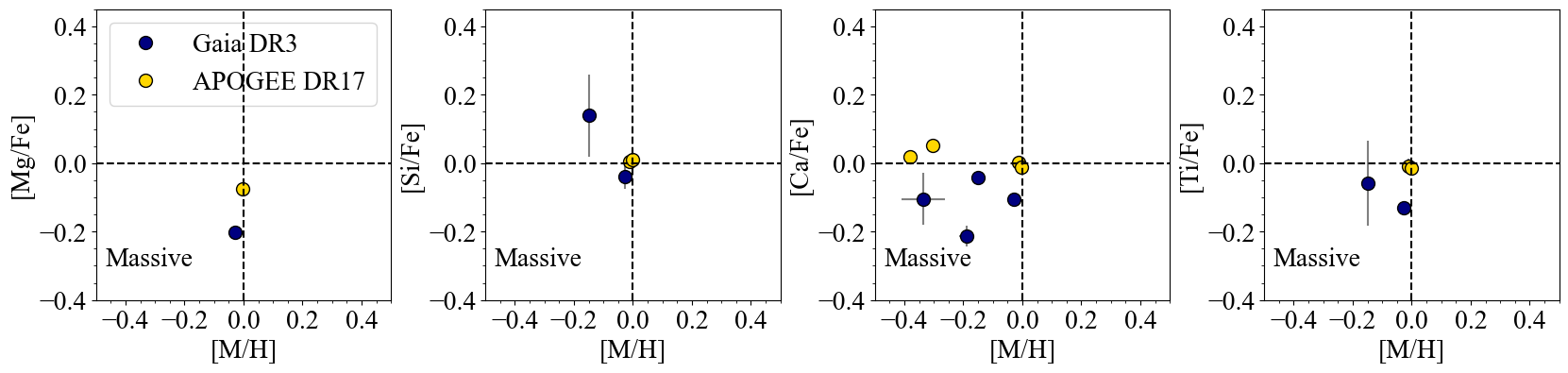}
\caption{   {\it Upper panels}: Comparison between the GSP-spec chemical abundances  ratios of  \gaia DR3   younger than 1 Gyr  analysed in this work  (blue points and associated errors) and  their counterparts in APOGEE DR17 (yellow points). {\it Lower panels}: As the upper panels, but for  the massive   \gaia DR3 stars  considered in this work.  }
\label{crossmatch_1gyr}
\end{centering}
\end{figure*}

\begin{figure}
\begin{centering}
\includegraphics[scale=0.45]{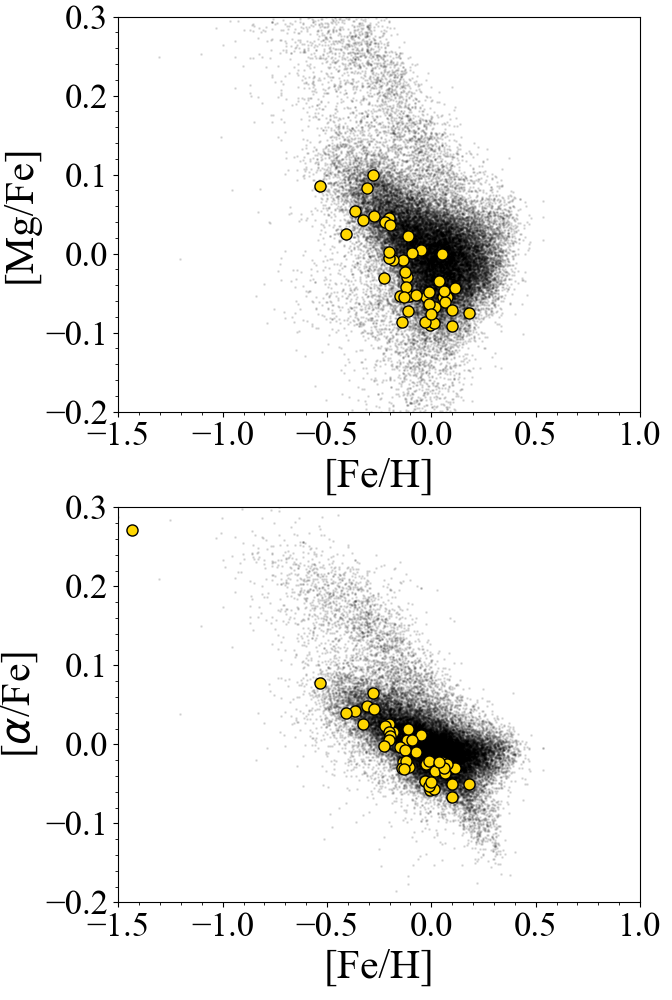}
\caption{  {\it Upper panel}: [Mg/Fe] versus [Fe/H] abundance ratios for APOGEE DR17 stars with Galactocentric distances  between 7 and 9 kpc  are reported with the black points. APOGEE DR17 stars which cross-match  with the population of massive stars presented by \citet{recioDR32022b}  characterised by chemical impoverishment with maximum vertical heights $|z_{max}|< 0.5$ kpc are indicated with the yellow points. {\it Lower panel}: As the upper one but for  [$\alpha$/Fe] versus [Fe/H]. 
}
\label{PVP}
\end{centering}
\end{figure}

\begin{figure}
\begin{centering}
\includegraphics[scale=0.45]{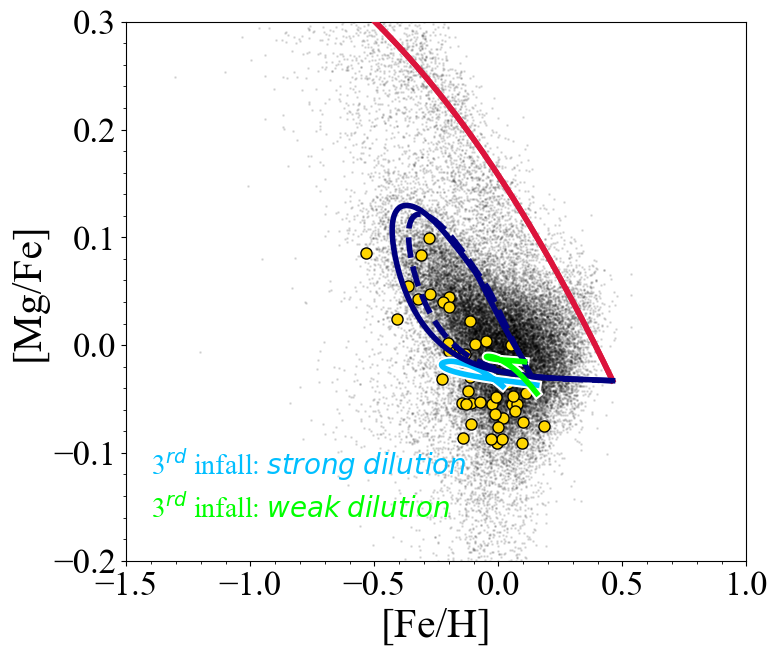}
\caption{ Comparison between the three-infall chemical evolution models presented in this study  and APOGEE DR17  data in the [Mg/Fe] versus [Fe/H] abundance ratios.  Symbols of the APOGEE DR17 stars are as in  the upper panel of Fig. \ref{PVP}. Line types and colours   of  the two chemical evolution  models  are  the ones indicated in Figs. \ref{si_sfr} (model with strong dilution created by  the third infall) and \ref{si_sfr_nodilu}  (model with a less massive third infall).
}
\label{models_apogee}
\end{centering}
\end{figure}

 In the left panel of Fig. \ref{kiel_massive}, we show the Kiel diagram of massive stars as selected  in the $T_{eff}$ and log g (calibrated values) space   by \citet{recioDR32022b} (see their Fig. 8) for the sub-samples considered in this work  for $\alpha$, Ca, Mg, Si and Ti. In this particular analysis, also for  the Mg sample, stars have been selected imposing the  'UpLim' flag  equal to 0 (best quality data).

In Fig. \ref{massive}, we compare the predicted  [X/Fe] versus [Fe/H] abundance ratios  for SSPs  younger than  1 Gyr (for different $\alpha$ elements) by the 'synthetic' model with the massive stars  reported in Fig. \ref{kiel_massive} 
with the calibrations  as proposed by \citet{recioDR32022a}. 
The only predictions that are not in agreement with the data are for Mg.  Although [Mg/Fe] ratios computed by our 'synthetic' model have larger values,   the  observed [M/H] is nicely traced  also in this case. 
 Following \citet{poggio2022}, in the right panel of Fig. \ref{kiel_massive},  we compare the  selected massive stars of  \gaia DR3 stars with the prediction from the PARSEC isochrones \citep{bressan2012,chen2014,chen2015,tang2014,pastorelli2019}  computed at  1 Gyr, 500 Myr, 300 Myr, 100 Myr and 30 Myr. These isochrones consider both solar metallicity [M/H]=0 dex and -0.25 dex because, as can be seen in Fig. \ref{massive},   the range in  metallicity spanned by the considered \gaia DR3 stars  $\in$ [-0.25, 0] dex. It is clear that the sample mostly contains young stars   with  ages  between 100 Myr and 300 Myr.

 An important test to validate  the young and chemical impoverished population  identified by the  \gaia DR3 is the comparison with the  high resolution spectroscopic surveys APOGEE DR17 \citep{apogeedr172022}.  In Fig. \ref{crossmatch_1gyr}, we show the stars in common between the sub-samples of \gaia DR3   and APOGEE DR17,  both for the massive stars  and  for objects younger  than 1 Gyr, respectively in the [X/Fe] versus [M/X] chemical space (where X= Mg, Si, Ca, Ti). 
Notwithstanding  some differences in the  [X/Fe] values - which can be explained by the differences in models/wavelength domain -   the metallicities  estimated  by both surveys  are in very good agreement. {\it It is important to note  that   only a few stars are  in common  between \gaia DR3 and  APOGEE DR17 in the  selected Galactic region ($R_g \in$ [8.1, 8.4] kpc, see Section \ref{s:gaia}).   As shown in Fig. \ref{massive},  \gaia DR3 shows a significant number of massive stars even in the
narrow region selected around the solar position.}

In addition, in Fig. \ref{PVP} we investigate in more details the massive young population introduced in Fig. 11 of      \citet{recioDR32022b} characterised by chemical impoverishment signatures  in [Ca/Fe] versus [M/H] ratios, showing the cross-match with APOGEE DR17.
Black points  indicate  [X/Fe] versus [Fe/H] (where X=Mg, $\alpha$) of APOGEE DR17 with Galactocentric distances in the range 7-9 kpc. With the  yellow  points we also draw the chemical ratios observed in   APOGEE DR17 for  the same stars as  in \citet{recioDR32022b}  which display chemical impoverishment in their Fig. 11 in the same Galactic region and   with maximum vertical heights $|z_{max}|< 0.5$ kpc.
From Fig. \ref{PVP}, it is very important for our analysis to conclude that  all the cross-matched stars  i) are part of the low-$\alpha$  sequence in APOGEE DR17 ii) the majority of them presents sub-solar values in metallicity. 
Finally, in Fig. \ref{models_apogee},  we compare  the three-infall chemical evolution models presented in this study  with  APOGEE DR17  data in the [Mg/Fe] versus [Fe/H] of Fig. \ref{PVP}.   The model which well reproduces GSP-Spec chemical abundances at different ages (model with strong dilution in the third infall phase  with parameters listed in Table \ref{tab_3I}) is also in better agreement with the APOGEE DR17 chemical abundances of  the cross-matched stars (yellow points).

 In order to further validate GPS-Spec chemical abundances, we compared  [Fe/H] values  for Cepheids taken from literature  with best \& medium quality GSP-Spec metallicities [M/H]   (calibrated values). A very good agreement is found, in fact  the average difference  [Fe/H]$_{\mbox{Lit.}}$-[M/H]$_{\mbox{GSP-Spec}}$ is -0.03 dex with $\sigma$= 0.11 dex (V. Ripepi, priv. comm.).

\begin{figure*}
\begin{centering}
\includegraphics[scale=0.22]{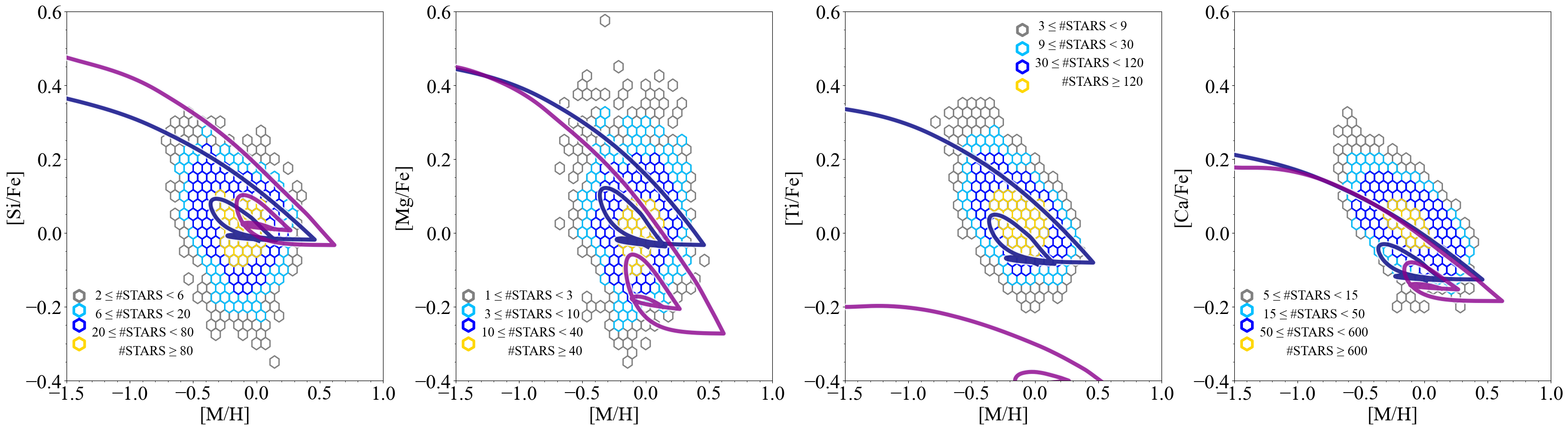}
\caption{  Observed \gaia DR3  [Si/Fe], [Mg/Fe], [Ti/Fe] and [Ca/Fe] 
abundance ratios versus [M/H] are reported with empty hexagons (the numbers of stars contained in different hexagons are indicated in the legends) are
compared with the proposed three-infall  chemical evolution model using different nucleosynthesis prescriptions.  Model predictions  (assuming parameters as reported in Table \ref{tab_3I})  adopting   the nucleosynthesis yields suggested by \citet{francois2004}  and \citet{romano2010} are reported with the blue and magenta lines, respectively.
}
\label{all_yield}
\end{centering}
\end{figure*}

\subsection{Results with \citet{romano2010} nucleosynthesis prescriptions}\label{romano_yields}
In this Section, we  present the three-infall model results using the nucleosynthesis prescriptions suggested by \citet{romano2010} (see Section \ref{romano_yields} for details) with the same parameters indicated as in Table \ref{tab_3I}.
In order to be consistent with \citet{romano2010} study, we    consider the IMF of \citet{kroupa1993}.
In  the different panels of Fig. \ref{all_yield}, we show the results for [Si/Fe], [Mg/Fe], [Ti/Fe] and [Ca/Fe] versus [M/H], respectively. We note that for Mg and Ti     the  predicted  abundance ratios  fall substantially  below  the observation data by more than 0.2 dex.
We stress, however, that modeling the galactic chemical evolution of Mg and Ti is challenging, even when adopting the most up-to-date nucleosynthesis prescriptions for massive stars. 
  For instance, \citet{prantzos2018} presented chemical evolution with metallicity-dependent weighted rotational velocities by \citet{limongi2018} but, the evolution of Mg and Ti substantially underestimate the observational data in the Milky Way. We want to stress that the largest uncertainties in the model predictions are still due to the adopted stellar nucleosynthesis prescriptions.

\section{Conclusions}\label{conc}

In this article, we have presented a new {\it three-infall} chemical  evolution model for the solar vicinity.   
Our main goal  was to extend the previous models presented  in \citet{spitoni2020} and \citet{spitoni2021}  in the light of the new constraints given by the star formation history (from  \gaia DR1 and \gaia DR2) and the RVS spectra abundances ratios  (\gaia DR3). 
Because these models were already able  to successfully reproduce data from high resolutions survey such as  APOKASC,  and APOGEE DR16, 
we kept unchanged the parameters proposed by these first studies  (see  Table \ref{tab_3I}).
Our main conclusions can be summarised as follows:
\begin{enumerate}

\item The  new {\it three-infall} chemical evolution model - characterised by infall time-scales $t_1=0.10$ Gyr, $t_2=4.11$ Gyr and $t_3$=0.15 Gyr,    and a strong chemical dilution during the third infall phase - nicely reproduces the main features of the abundance ratio [X/Fe] versus [M/H]  (X=Mg, Si, Ca, Ti, $\alpha$) of \gaia DR3 stars  in different age bins for the considered $\alpha$ elements.

\item The most recent gas  infall - which started $\sim$ 2.7 Gyr ago -  allows us to predict well the \gaia DR3 young population  which has suffered a recent chemical impoverishment. We have shown that that the classical two-infall model   is not capable to predict this young population centered
at undersolar [M/H] values because too many  stars at 
higher metallicity are formed.

\item Our model also  reproduces important observational constraints for the chemical evolution of the Galactic disc  such as  the present-day star formation rate, the  stellar and gas surface densities, Type II and Ia SN rates,  and the infall rate of the accreted gas. In addition, the  proposed three-infall model is able to reproduce the solar photospheric abundance values of \citet{grevesse2007}.

\item   The stars  identified by \citet{recioDR32022b}  as massive objects  - with the above mentioned signature of chemical impoverishment - are also part of  low-$\alpha$ sequence considering the associated APOGEE DR17 chemical abundances.
\item The proposed three-infall model - with a strong dilution in the chemical abundance ratios caused by the third infall event -  is capable to reproduce reasonably well also APOGEE DR17 data.

\item  Future \gaia data releases, as well as other surveys, might revise the properties and significance of the young population of metal-impoverished stars found by \citet{recioDR32022b}.  However, {\it the three-infall model presented here, constrained by the recent local SFH derived from Gaia data, would still remain a viable scenario for the local disc formation.} It could be necessary, though, to assume a higher/smaller amount of accreted gas.

\end{enumerate}
The above mentioned results suggest that  the distribution of the stars in high-$\alpha$ and low-$\alpha$ sequences is strictly linked to different star formation regimes over the Galaxy’s life.
As already pointed out by \citet{recioDR32022b}, 
 stellar migration is not capable to  explain the presence of the  significant young stellar  population with an impoverished chemical composition because  inwards migration from outer parts  (the one which will favor lower metallicities) should not dominate due to the decreasing stellar density with the Galactocentric distance. In conclusion, we suggest that the main reason for  this peculiar young stellar population is the chemical dilution from a recent gas infall episode.

\section*{Acknowledgement}

E. Spitoni and A. Recio-Blanco received funding from the European Union’s Horizon 2020 research and innovation program under SPACE-H2020 grant agreement number 101004214 (EXPLORE project). 
This project has received funding from the European Union's Horizon 2020 research and innovation programme under the Marie Sklodowska-Curie grant agreement N. 101063193.
This work has made use of data from the European Space Agency (ESA) mission
\gaia (\url{https://www.cosmos.esa.int/gaia}), processed by the \gaia
Data Processing and Analysis Consortium (DPAC,
\url{https://www.cosmos.esa.int/web/gaia/dpac/consortium}). Funding for the DPAC
has been provided by national institutions, in particular the institutions
participating in the \gaia Multilateral Agreement.

\bibliographystyle{aa} 
\bibliography{disk}

\end{document}